\documentclass[prb,preprint,showpacs,preprintnumbers,amsmath,amssymb,aps]{revtex4-1}
\usepackage{graphicx}
\usepackage{dcolumn}
\usepackage{bm}
\usepackage{color}
\def\be{\begin{equation}}       \def\ee{\end{equation}}
\def\bea{\begin{eqnarray}}      \def\eea{\end{eqnarray}}
\def\bes{\begin{subequations}}  \def\ees{\end{subequations}}

\def\dag{\dagger}
\def\non{\nonumber}

\def\k{_{\bf k}}
\def\km{_{-\bf k}}

\begin{document}

\title{Spin dynamics of antiferromagnetically coupled ferromagnetic bilayers -- the case of Cr$_2$WO$_6$ and Cr$_2$MoO$_6$}
\author{Kingshuk Majumdar}
\email{majumdak@gvsu.edu}
\affiliation{Department of Physics, Grand Valley State University, Allendale, 
Michigan 49401, USA}
\author{Subhendra D. Mahanti}
\email{mahanti@pa.msu.edu}
\affiliation{Department of Physics and Astronomy, Michigan State University, East Lansing, 
Michigan 48824, USA}

\date{\today}

\begin{abstract}
\label{abstract}
Recent inelastic neutron diffraction measurements on Cr$_2$(Te, W, Mo)O$_6$ have revealed 
that these systems consist of bilayers of spin-3/2 Cr$^{3+}$ ions with strong antiferromagnetic inter-bilayer coupling and 
tuneable intra-bilayer coupling from ferro (for W and Mo) to antiferro (for Te). These measurements have determined
the ground state spin structure and the values of sublattice magnetization, which shows significant reduction of sublattice 
magnetization from the atomic spin value of 3.0$\mu_B$ for Cr$^{3+}$ atoms. In an earlier paper we theoretically
investigated the low temperature spin dynamics of Cr$_2$TeO$_6$ bilayer system where both the intra and inter-bilayer couplings are 
antiferromagnetic. In this paper we investigate Cr$_2$WO$_6$ and Cr$_2$MoO$_6$ systems where intra-bilayer exchange couplings are 
ferromagnetic but the inter-bilayer exchange couplings are antiferromagnetic. We obtain the magnon dispersion, sublattice magnetization, 
two-magnon density of states, longitudinal spin-spin correlation function, and its powder average and compare 
the results for these systems with results for Cr$_2$TeO$_6$. 
\end{abstract}

\pacs{71.15.Mb, 75.10.Jm, 75.25.-j, 75.30.Et, 75.40.Mg, 75.50.Ee, 73.43.Nq}

\maketitle

\section{\label{sec:Intro}Introduction and Formalism}

Exploring the dynamics of quantum spins with competing interactions and geometrical frustration has been one of the most exciting 
areas of theoretical and experimental research over last several decades.~\cite{anderson,harris71,diep,mila,rastelli,majumdar10,
majumdar11a, majumdar11b,majumdar12,majumdar13} 
A subset of this research is understanding the physics 
of interacting quantum spin dimers (QSD), where the intra-dimer interaction is antiferromagnetic (strength $J$).~\cite{subir,zhu14,mahanti91}
By tuning the geometry and the strength of inter-dimer coupling ($j$), the system can go from strongly fluctuating zero-dimensional to 
quasi $n$ ($n=1,2,3$) dimensional system, accompanied by dramatic changes in spin dynamics.~\cite{zhu14,zhu15}

Some of the interesting observations for the ground state of interacting quantum spins (IQS) are: absence of long range order (LRO) 
and long range quantum spin entanglement i.e. a liquid like structure (for example, Haldane state for 1D chains with integer spins, 
Luttinger Fermi-liquid for half-integer spins) in 1D even at $T=0$K or dramatic reduction in the LRO moment in 2D due to 
quantum fluctuations.  The excitations also span a broad range, from spinons to triplons to magnons. These theoretical 
developments have lead to the synthesis of many interesting insulating magnetic systems where the spin dimensionality, space 
dimensionality, inter-spin coupling can be tuned. Experimental studies in these systems have deepened our fundamental 
understanding of the physics of IQS systems.~\cite{diep,mila,rastelli,subir,ronnow01,chris04,chris07}

In a particular class of IQS, one of present interest, the system consists of quantum spin dimers (QSDs). Depending on the nature 
of the super exchange coupling between the localized magnetic moments, the dominant interaction is the intra-dimer coupling $J$. 
In this case, the magnetic centers are QSDs with antiferromagnetic $J$ weakly interacting with each other through $j$. 
If, on the other hand, the interaction between the QSDs is stronger than $J$, then the system can be thought of as 2$D$ ferro- 
or antiferro-magnetic sheets with antiferromagnetic inter-sheet coupling. In fact, by manipulating local chemistry one can 
tune this coupling from F to AF, going through effectively non-interacting ($j=0$) QSDs.~\cite{zhu14,zhu15}

The focus of this paper is to explore the effect of changing the sign and strength of $j$ on the ground and excited states 
using the example of a Cr based system, Cr$_2$XO$_6$ (X= Te, W, Mo). On theoretical ground one expects the system to undergo a 
quantum phase transition from a quantum disordered state to a state with LRO as $|j|$ is increased. The latter state supports 
magnon excitations. If one is not too far from the critical region then the resulting soft magnons reduce the LRO moment. 
How the magnon dispersion and reduction in the moment depends on the sign of $j$ are interesting questions that we explore 
in this paper. It is experimentally found that in Cr$_2$TeO$_6$, $j$ is anti-ferromagnetic whereas in Cr$_2$WO$_6$ and 
Cr$_2$MoO$_6$, $j$ is 
ferromagnetic.~\cite{zhu15} This unusual observation was explained by {\it ab initio} density functional theory based calculations of 
different magnetically ordered states in these compounds, and was ascribed to the presence of low energy unoccupied 
$d$-states in W and Mo, an  idea similar to $d$-zeroness in ferroelectricity.~\cite{nicola} In spite of the fact 
that $J$ is the dominant exchange coupling, due to sufficiently large $|j|$  and the number of inter-dimer bonds, these systems 
show LRO and the excitations are magnon-like. In addition to magnon modes there is strong experimental evidence of 
Higgs-like amplitude modes, a characteristics of interacting QSDs.~\cite{zhu15} Here we will discuss only the 
magnon-like excitations. The case of antiferromagnetic $j$ has been extensively discussed in an earlier paper by us.~\cite{majumdar18a} 
In this paper, we will discuss the results for the ferromagnetic case briefly focusing on the similarities and 
differences between the two cases.

\begin{figure}[httb]
\centering
(a)
\includegraphics[width=2.3in,clip]{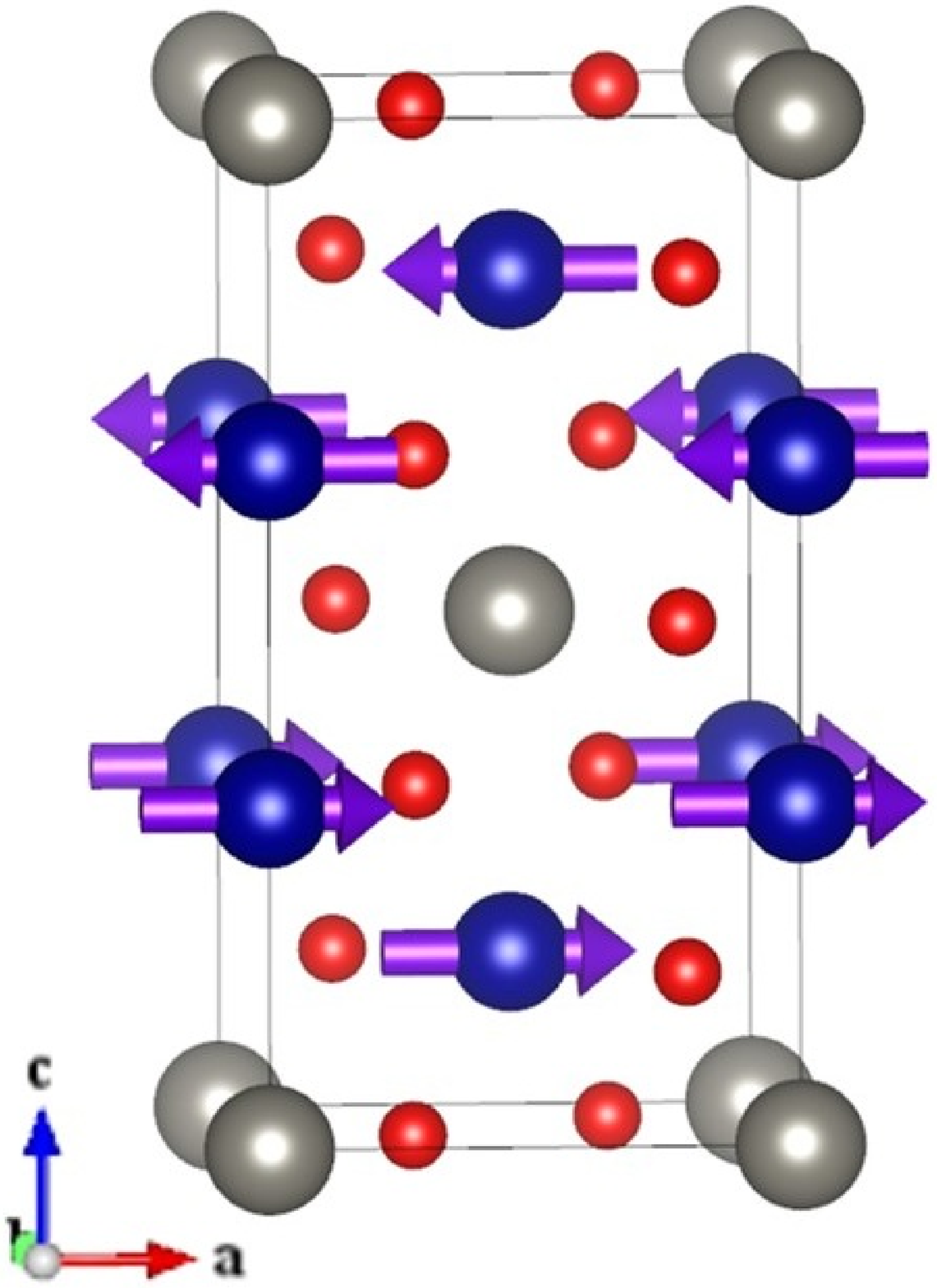}
\qquad 
(b)
\includegraphics[width=3.0in,clip]{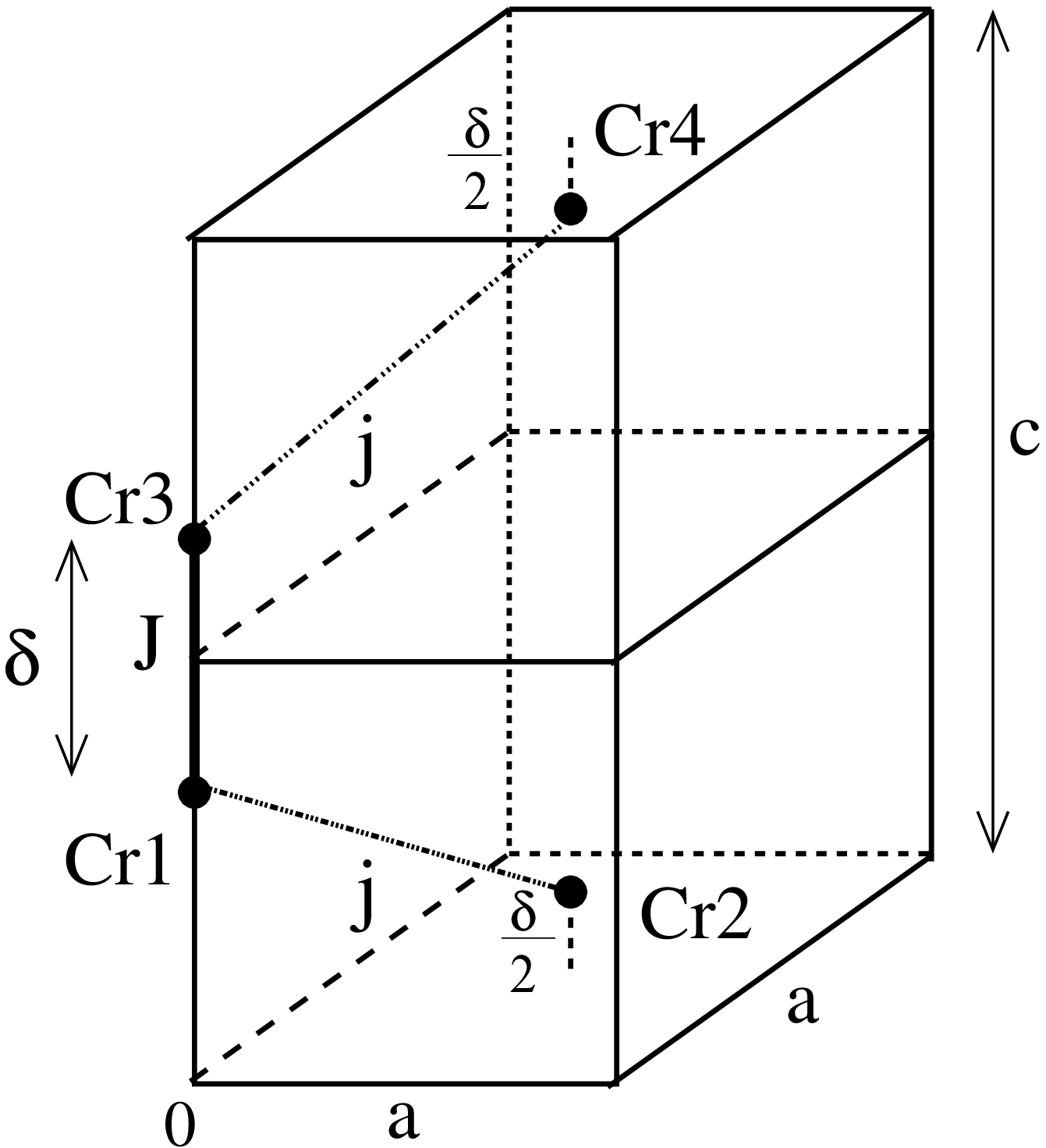}
\caption{\label{fig:CrMWstruc1} (Color online)(a) Schematic of the bilayer crystal structure and magnetic ordering of Cr$_2$(Mo, W)O$_6$.
Each Cr$^{3+}$ (blue spheres) bilayer is separated by a W/Mo (grey spheres) layer.~\cite{zhu14,zhu15} The oxygen (O) atoms are shown by the red spheres. (b) 
Positions of four chromium spins in the tetragonal unit cell of dimensions
$(a,a,c)$ are shown. The coordinates of the spins are given in Ref.\onlinecite{majumdar18a}.}
\end{figure}

In Fig.~\ref{fig:CrMWstruc1} we show the ground state spin ordering in 
Cr$_2$(X=W, Mo)O$_6$.~\cite{kunn68} One has two bilayers (perpendicular to the $z$-axis) in the 
tetragonal unit cell $(a,a,c)$ and four Cr 
spins/unit cell. The experimental unit cell parameters for Cr$_2$WO$_6$ are 
$a=4.583$\AA,\; $c=8.853$\AA\; and $a=4.587$\AA, $c=8.811$\AA\; for Cr$_2$MoO$_6$ at $T=4$K.~\cite{zhu14} 
The Cr-O-Cr bond angles and bond lengths of both of these compounds are similar due to the similar ionic radii of Mo$^{6+}$ and W$^{6+}$.
The shortest distance between the inter-bilayer (NN) Cr atoms i.e. Cr1 and 
Cr3 is $\delta \sim 3.00$\AA\;$\approx c/3$, whereas the distance between 
intra-bilayer NN Cr atoms (Cr1 and Cr2 or Cr3 and Cr4) is $\sim 3.80$\AA. One bilayer contains Cr1 and Cr2 spins and the other 
contains Cr3 and Cr4 spins. The  
inter-bilayer AF coupling $J$ comes through Cr1-Cr3 and Cr2-Cr4 dimers. The NN intra-bilayer ferromagnetic coupling $j$ is 
between Cr3-Cr4 and Cr1-Cr2.
Estimates of exchange parameters from high temperature thermodynamic measurements~\cite{drillon79} indicate 
that $|j|, |j^\prime| << |J|$ -- so these systems can be regarded as weakly 
interacting quantum dimers.  

In this paper we calculate
magnon dispersion, sublattice magnetization, two-magnon density of states, longitudinal spin-spin correlation function, 
and it's  powder average using linear spin-wave theory.~\cite{majumdar10, majumdar12} Our current work is for a 
completely different class of systems with different ground state spin configuration than our 
recently published work on Cr$_2$TeO$_6$.~\cite{majumdar18a} In this paper we briefly provide the theoretical formalism in Appendix~\ref{formalism} 
and present only the 
relevant equations and results pertinent to the current systems.  

\subsection{Magnon Dispersion and Sublattice Magnetization}

The Heisenberg Hamiltonian of systems with 
F intra- and AF inter-bilayer couplings $j, j^\prime$, and $J$ ($j,\;j^\prime,\;J>0$) has the form
\be
{\cal H}={\cal H}_{\rm NN}+{\cal H}_{\rm NNN},
\ee
with
\begin{subequations}
\bea
{\cal H}_{\rm NN} &=& -j\sum_{n=1}^{N_z} \sum_{\langle i,j\rangle} \Big[ {\bf S}_{in}^{(1)A}\cdot {\bf S}_{jn}^{(2)A}
      +{\bf S}_{in}^{(3)B}\cdot {\bf S}_{jn}^{(4)B}\Big]\non \\
&+&J\sum_{n=1}^{N_z} \sum_{i} \Big[ {\bf S}_{in}^{(1)A}\cdot {\bf S}_{in}^{(3)B}
      +\frac 1{2}\{ {\bf S}_{in}^{(2)A}\cdot {\bf S}_{in-1}^{(4)B}+{\bf S}_{in}^{(4)B}\cdot {\bf S}_{in+1}^{(2)A}\}\Big],\\
{\cal H}_{\rm NNN}&=&-j^\prime \sum_{n=1}^{N_z}\sum_{\langle \langle i,j\rangle \rangle}\Big[{\bf S}_{in}^{(1)A}\cdot {\bf S}_{jn}^{(1)A}
+{\bf S}_{in}^{(2)A}\cdot {\bf S}_{jn}^{(2)A}+{\bf S}_{in}^{(3)B}\cdot {\bf S}_{jn}^{(3)B}+{\bf S}_{in}^{(4)B}\cdot {\bf S}_{jn}^{(4)B}\Big].     
\eea
\label{ham}
\end{subequations}
After Holstein-Primakoff transformation~\cite{HP} and Fourier transform, the quadratic part of the 
Hamiltonian represented in terms of interacting bosons 
$a$ and $b$ takes the form (the details are shown in Appendix~\ref{formalism}):
\begin{eqnarray}
{\cal H}_0 &=& jS(4+\eta)\sum\k \kappa\k\Big[\Big(a\k^{{(1)}\dag} a\k^{(1)}+a\k^{{(2)}\dag} a\k^{(2)}+b\km^{{(3)}\dag} b\km^{(3)}
+b\km^{{(4)}\dag} b\km^{(4)}\Big) \nonumber \\
&-& \gamma_{1 \bf k}\Big(a^{(1)}\k a^{(2)\dag}\k+b^{(3)}\km b^{(4)\dag}\km \Big)- 
\gamma_{1 \bf k}^*\Big(a^{(2)}\k a^{(1)\dag}\k+b^{(4)}\km b^{(3)\dag}\km \Big) \nonumber \\
&+& \gamma_{2 \bf k}\Big(a^{(2)}\k b^{(4)}\km+a^{(1)\dag}\k b^{(3)\dag}\km \Big)+ 
\gamma_{2 \bf k}\Big(a^{(1)}\k b^{(3)}\km+a^{(2)\dag}\k b^{(4)\dag}\km \Big)
\Big],
\label{quadH}
\end{eqnarray}
where, 
\bea
 \gamma_{1 \bf k}&=& \frac 4{4+\eta} \frac {e^{ik_z c/2}\cos (k_xa/2)\cos (k_y a/2)}{1+\gamma_{3\bf k}}, \non \\
 \gamma_{2 \bf k} &=& \frac {\eta}{4+\eta}\frac 1{1+\gamma_{3\bf k}},\non \\
 \gamma_{3 \bf k}&=& \frac {4\eta^\prime}{4+\eta}[1-\frac 1{2}(\cos(k_x a)+\cos (k_y a))], \non \\
 \kappa\k &=& 1+\gamma_{3\bf k}.
\eea
Above, $\eta=J/j$ and $\eta^\prime=j^\prime/j$.
${\cal H}_0$ in Eq.~\eqref{quadH} can be 
succinctly written as ${\cal H}_0 = {\cal H}_1 \oplus {\cal H}_1^{\rm T}$ where ${\cal H}_1^{\rm T}$ is the transpose of ${\cal H}_1$.
In the Fourier transformed basis $X_{\bf k}= (a\k^{(1)}\; a\k^{(2)}\; b\km^{(4)\dag}\; b\km^{(3)\dag})^T$ we write ${\cal H}_1$ as
\be
{\cal H}_1=jS(4+\eta)\sum_{\bf k}\kappa\k X_{\bf k}^\dag {\cal H}_{1\bf k}^\prime X_{\bf k}
\ee
with 
\be
{\cal H}_{1\bf k}^\prime= 
\begin{bmatrix}A\k & B\k & 0 & C\k \\ 
B\k^\star & A\k & C\k & 0 \\
0 & C\k & A\k & B\k \\ 
C\k & 0 & B\k^\star & A\k \end{bmatrix}
\ee
and $A\k=1, B\k=-\gamma_{1\bf k}^\ast, C\k=\gamma_{2 {\bf k}}$. 
Next, we diagonalize ${\cal H}_1$ by transforming the 
operators $a_{\bf k}$ and $b_{\bf k}$ to magnon operators 
$\alpha_{\bf k}$ and $\beta_{\bf k}$ using the following generalized Bogoliubov (BG) transformations~\cite{bogoliubov, colpa, wheeler, trinanjan}:
\bea
\begin{pmatrix}a\k^{(1)} \\ a\k^{(2)}\\ b\km^{(4)\dagger}\\ b\km^{(3)\dagger} \end{pmatrix}
=\begin{bmatrix}\ell_{1\bf k} &  \ell_{1\bf k}^\prime & m_{1\bf k} & m_{1\bf k}^\prime \\ 
\ell_{2\bf k}^\prime & \ell_{2\bf k} & m_{2\bf k}^\prime & m_{2\bf k} \\
m_{1\bf k} & m_{1\bf k}^\prime & \ell_{1\bf k} & \ell_{1\bf k}^\prime \\ 
m_{2\bf k}^\prime & m_{2\bf k} & \ell_{2\bf k}^\prime & \ell_{2\bf k} \end{bmatrix}
\begin{pmatrix}\alpha \k^{(1)} \\ \alpha \k^{(2)}\\ \beta\km^{(1)\dag}\\ \beta\km^{(2)\dag} \end{pmatrix}.
\eea
The elements of the transformation matrix $\ell_{(1,2)\bf k},\ell_{(1,2)\bf k}^\prime,m_{(1,2)\bf k},m_{(1,2)\bf k}^\prime$ are given in 
Appendix~\ref{lmcoeffs}.

The  quadratic Hamiltonian after diagonalization becomes:
\bea
\label{diagH}
{\cal H}_1 &=& jS(4+\eta)\sum\k \kappa\k \Big\{ \omega_{\bf k}^{(1)}\Big[\alpha\k^{(1)\dag}\alpha\k^{(1)}+ \beta\km^{(1)\dag}\beta\km^{(1)}\Big] 
+ \omega_{\bf k}^{(2)}\Big[\alpha\k^{(2)\dag}\alpha\k^{(2)}+ \beta\km^{(2)\dag}\beta\km^{(2)}\Big]\Big\}\non \\
&-&jS(4+\eta)\sum\k \kappa\k\Big[\omega^{(1)}_{\bf k}+\omega^{(2)}_{\bf k}-2\Big].
\eea
${\cal H}_1^T$ has the same structure as ${\cal H}_1$. The two roots in Eq.~\eqref{diagH} are:~\cite{wheeler, trinanjan} 
\be
\omega_{\bf k}^{(1,2)}=\Big[A\k^2+|B\k|^2 -C\k^2 \mp \sqrt{4A\k^2\vert B\k \vert^2
-C\k^2|B\k^\star -B\k|^2}\;\Big]^{1/2}.
\ee
For our case, the eigenvalues for the $\alpha$ and $\beta$ magnon branches (a low energy acoustic branch and a high
energy optic branch) simplify to
\be
\omega_{\bf k}^{(1,2)}=\Big[1+\vert \gamma_{1 \bf k }\vert^2-\gamma_{2 \bf k }^2 
\mp \sqrt{4\vert \gamma_{1 \bf k} \vert^2-\gamma_{2\bf k}^2\vert \gamma_{1\bf k}-\gamma_{1\bf k}^*\vert^2}\;\Big]^{1/2}
\ee
and the quasiparticle energies $E_{\bf k}^{(1,2)}$ for these magnons are given by: 
\be
E_{\bf k}^{(1,2)}=jS(4+\eta)\kappa\k \omega\k^{(1,2)}. \label{dispersion}
\ee
The second term in Eq.~\eqref{diagH} is the quantum-zero point energy, which contributes to 
the ground state
energy. 
In order to understand the physical origin of the two modes with frequencies $\omega_{\bf k}^{(1)}$ and $\omega_{\bf k}^{(2)}$, each 
two-fold degenerate (for the full quadratic ${\cal H}_0$), we start from the limit when the inter-bilayer coupling $J=0$ 
and then introduce nonzero $J$. When $J=0$, we have two decoupled ferromagnetic bilayers. In anticipation 
of antiferromagnetic $J$, we denote one bilayer spins ``up'' ($\alpha$-magnons) and the other bilayer spins ``down'' ($\beta$-magnons). They 
are of course degenerate, each  with two modes of frequencies  $\omega_{\bf k}^{(1)}$ and $\omega_{\bf k}^{(2)}$. These two modes 
arise as the unit cell contains two spins of each orientation. For the ferromagnetic ordering we could have chosen a smaller unit 
cell with one spin/unit cell and one would have obtained one ferromagnetic magnon branch. When mapped on to the smaller BZ 
associated with larger unit cell (two spins/unit cell) we get two branches. For simplicity, we can refer to these two 
branches as acoustic and optic branches in analogy with phonons. Thus in the limit $J=0$, we have a two-fold degenerate 
acoustic branch (one $\alpha$ and one $\beta$) and a two-fold degenerate optic branch (one $\alpha$ and one $\beta$). When we turn 
on $J$, the degenerate $\alpha$ and $\beta$ branches mix and give rise to new $\alpha$ and $\beta$ branches which preserve their 
double degeneracy because of time-reversal symmetry, similar to the case of magnons in a simple antiferromagnet. 

The normalized sublattice magnetization $m_s=M_s/M_0$ (where $M_0=g\mu_B$) for the A-sublattice can be expressed as 
\be m_s=S-\delta S, 
\label{MagF}
\ee
where,
\be 
\delta S =\frac 1{N}\sum\k \langle a^{(1) \dag}\k a^{(1)}\k \rangle
= \frac 1{N}\sum\k \Big[\vert m_{1\bf k}\vert^2+\vert m_{1\bf k}^\prime\vert^2\Big].
\label{dS-F}
\ee
$\delta S$ corresponds to the reduction of magnetization within linear spin-wave theory (LSWT) and the summation over ${\bf k}$ goes over the entire
Brillouin zone corresponding to the tetragonal unit cell $(a,a,c)$. The Bogoliubov coefficients $m_{1\bf k}$ and $m_{1\bf k}^\prime$
in Eq.~\eqref{dS-F} are given in Appendix~\ref{lmcoeffs}.

\subsection{Two-magnon density of states (TM-DOS) and Longitudinal spin-spin correlation function (LSSCF)}
\begin{figure}[httb]
\centering
\includegraphics[width=3.1in,clip]{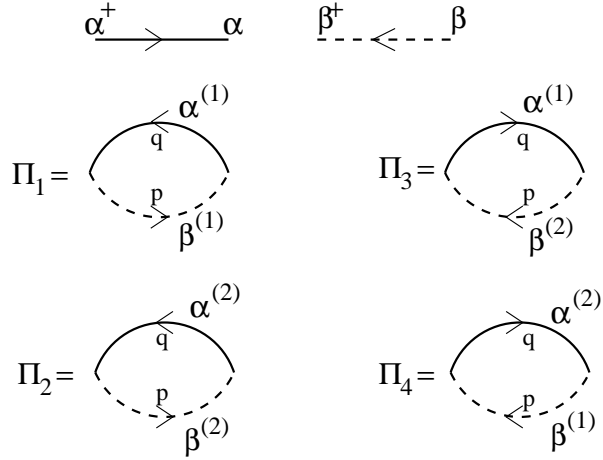}
\qquad 
\caption{\label{fig:FeynPi} Green's function propagators for $\alpha$ and $\beta$ magnons (1 and 2) are shown by solid and dashed lines
respectively. The Feynman diagram for the four time-ordered Green's functions $\Pi_{i=1\cdots 4}(\omega)$
that contribute to the longitudinal spin-spin correlation function are shown.}
\end{figure}

TM-DOS associated with the four magnon branches ($i,j=1,2$) are given as:
\be
{\rm DOS}_{ij}({\bf k}, \omega) = \sum_{\bf {p}}\delta (\omega-\omega^{(i)}_{\bf p}-\omega^{(j)}_{\bf k+p}). \label{dos}
\ee
DOS$_{11}$, DOS$_{22}$ are the intra-branch and DOS$_{12}$, DOS$_{21}$ 
are the inter-branch density of states. Longitudinal spin-spin correlation function ${\cal L}_s({\bf k}, \omega)$ is the sum of the 
weighted $\delta$-functions arising from these four density of states. LSSCF is defined as
\be
{\cal L}_s({\bf k}, t)=\langle S_z ({\bf k}, t) S_z(-{\bf k}, 0) \rangle,
\label{lss}
\ee
with
\be
S_z ({\bf k})=\frac 1{\sqrt {4N}}\sum_{i\mu} S_z^{i\mu} e^{-i{\bf k} \cdot ({{\bf R}_i}+\tau_\mu)}.
\ee
Here ${\bf R}_i$ is the position vector of the $i$-th unit cell and ${\bf \tau}_\mu$ are the positions of the four Cr-atoms in the unit cell. 
The position of the Cr-atoms are respectively: 
Cr1: $\tau_1=(0,0, c/2-\delta/2)$, Cr2: $\tau_2=(a/2,a/2,\delta/2)$,
Cr3: $\tau_3=(0,0, c/2+\delta/2)$, and Cr4: $\tau_2=(a/2,a/2,c-\delta/2)$ [See Fig.~\ref{fig:CrMWstruc1}].
Experimentally measured quantity is the Fourier transform of the time-dependent spin-correlation function
${\cal L}_s({\bf k},t)$
\be
{\cal L}_s({\bf k}, \omega)=\int_{-\infty}^{\infty} \frac {dt}{2\pi} {\cal L}_s({\bf k}, t)e^{-i\omega t}.
\ee
where spins for each of the sublattices $1, 2, 3, 4$ after Fourier transform become:
\begin{subequations}
\bea
S_z^{(\mu=1,2)}({\bf k}) &=& \sqrt{4N}S\delta({\bf k}=0)-\frac 1{\sqrt {4N}}\sum_{{\bf p,q}}\delta ({\bf k}+{\bf p}-{\bf q}) f_{\mu \bf k}
a_{\bf p}^{(\mu)\dagger}a_{\bf q}^{(\mu)}, \\
S_z^{(\mu=3,4)}({\bf k}) &=& -\sqrt{4N}S\delta({\bf k}=0)+\frac 1{\sqrt {4N}}\sum_{{\bf p,q}}\delta ({\bf k}+{\bf p}-{\bf q}) 
f_{\mu \bf k}b_{-{\bf q}}^{(\mu)\dagger}b_{-\bf p}^{(\mu)}.
\eea
\end{subequations}
$f_{\mu {\bf k}}=e^{-i{\bf k \cdot \tau_\mu}}$ takes into account the relative phases of the different magntic atoms inside 
the unit cell. The total spin can now be written as:
\be
S_z({\bf k})=-\frac 1{\sqrt{4N}}\sum_{{\bf p,q}}\delta ({\bf k}+{\bf p}-{\bf q}) 
\Big\{ [f_{1\bf k} a_{\bf p}^{(1)\dagger}a_{\bf q}^{(1)}+f_{2\bf k}a_{\bf p}^{(2)\dagger}a_{\bf q}^{(2)}]
-[f_{3\bf k}b_{-{\bf q}}^{(3)\dagger}b_{-\bf p}^{(3)}+f_{4\bf k}b_{-{\bf q}}^{(4)\dagger}b_{-\bf p}^{(4)}] \Big\}.
\ee
Using BG transformations we express $S_z({\bf k})$ in terms of the magnon operators $\alpha$ and $\beta$. The result is shown in the 
Appendix~\ref{SScorr}. There are $16 \times 16$ time-ordered Green's functions that arise from Eq.~\eqref{lss}, of which only four 
shown in Fig.~\ref{fig:FeynPi} contribute to LSSCF. These are defined in Ref.~\onlinecite{majumdar18a}.
The correlation function ${\cal L}_s ({\bf k}, \omega)$ takes the following form:
\bea
{\cal L}_s({\bf k}, \omega) &=& \frac 1{4N}\Big[\sum_{\bf p} \delta (\omega-\omega^{(1)}_{\bf p}-\omega^{(1)}_{\bf p+k})
\vert {\cal D}^{11}_{\bf k, k+p} \vert^2 
+\sum_{\bf p} \delta (\omega-\omega^{(2)}_{\bf p}-\omega^{(2)}_{\bf p+k})
\vert  {\cal D}^{22}_{\bf k, k+p} \vert^2 \non \\
&+&\sum_{\bf p} \delta (\omega-\omega^{(2)}_{\bf p}-\omega^{(1)}_{\bf p+k})
\vert {\cal D}^{21}_{\bf k, k+p} \vert^2 
+\sum_{\bf p} \delta (\omega-\omega^{(1)}_{\bf p}-\omega^{(2)}_{\bf p+k})
\vert  {\cal D}^{12}_{\bf k, k+p} \vert^2\Big],  \label{spin-corr}
\eea
where the weights $D^{ij}_{{\bf k, k+p}}$ are defined as,
\begin{subequations}
\bea
{\cal D}^{11}_{\bf k, k+p} &=& [f_{1\bf k}\ell_{1 \bf p}^\ast m_{1 {\bf p+k}}+f_{2\bf k}\ell_{2 \bf p}^{\prime \ast}m_{2 {\bf p+k}}^\prime] 
-[f_{3\bf k}\ell_{2 {\bf p+k}}^\prime m_{2 \bf p}^{\prime \ast}+f_{4\bf k}\ell_{1{\bf p+k}}m_{1 {\bf p}}^\ast ], \\
{\cal D}^{22}_{\bf k, k+p}&=& [f_{1\bf k}\ell_{1 \bf p}^{\prime \ast} m_{1 {\bf p+k}}^\prime +f_{2\bf k}\ell_{2 \bf p}^{\ast}m_{2 {\bf p+k}}]
-[f_{3\bf k}\ell_{2 {\bf p+k}} m_{2 \bf p}^{\ast}+f_{4\bf k}\ell_{1{\bf p+k}}^\prime m_{1 {\bf p}}^\ast ], \\
{\cal D}^{21}_{\bf k, k+p} &=& [f_{1\bf k}\ell_{1 \bf p}^{\ast} m_{1 {\bf p+k}}^\prime +f_{2\bf k}\ell_{2 \bf p}^{\prime \ast}m_{2 {\bf p+k}}]
-[ f_{3\bf k}\ell_{2 {\bf p+k}} m_{2 \bf p}^{\prime \ast}+f_{4\bf k}\ell_{1{\bf p+k}}^\prime m_{1 {\bf p}}^\ast ], \\
{\cal D}^{12}_{\bf k, k+p} &=& [f_{1\bf k}\ell_{1 \bf p}^{\prime \ast} m_{1 {\bf p+k}}+f_{2\bf k}\ell_{2 \bf p}^{\ast}m_{2 {\bf p+k}}^\prime]
-[f_{3\bf k}\ell_{2 {\bf p+k}}^\prime m_{2 \bf p}^{\ast}+f_{4\bf k}\ell_{1{\bf p+k}} m_{1 {\bf p}}^{\prime \ast}].
\eea
\end{subequations}
Smooth TM-DOS and ${\cal L}_s({\bf k}, \omega)$ are obtained 
by replacing the $\delta$-function by a Gaussian with a constant broadening width $\sigma$:
\be
\sum_{\bf {p}}\delta (\omega-\omega_{\bf p})\rightarrow \sum_{\bf {p}} \frac 1{\sqrt{2\pi \sigma^2}}e^{-(\omega-\omega_{\bf p})^2/2\sigma^2}.
\label{delta-fnc}
\ee
This Gaussian function accounts for three purposes: (1) finite experimental resolution, (2) experimental uncertainty in the 
determination of the continua, and (3) finite life time of the measured excitations induced by finite temperature and/or by disorder in the
sample.~\cite{powalski}  
The powder average of the longitudinal spin-spin correlation function is obtained by averaging over the angles $\theta$ and $\phi$
for a given value of $Q$:
\be
\langle {\cal L}_s (Q,\omega) \rangle = \frac 1{4\pi}\int_0^{2\pi} d\phi \int_0^{\pi} d\theta \sin \theta \;{\cal L}_s({\bf k}, \omega).
\label{powavg}
\ee
\section{\label{sec:results}Results}

\subsection{Magnon Energy Dispersion}
\begin{figure}[httb]
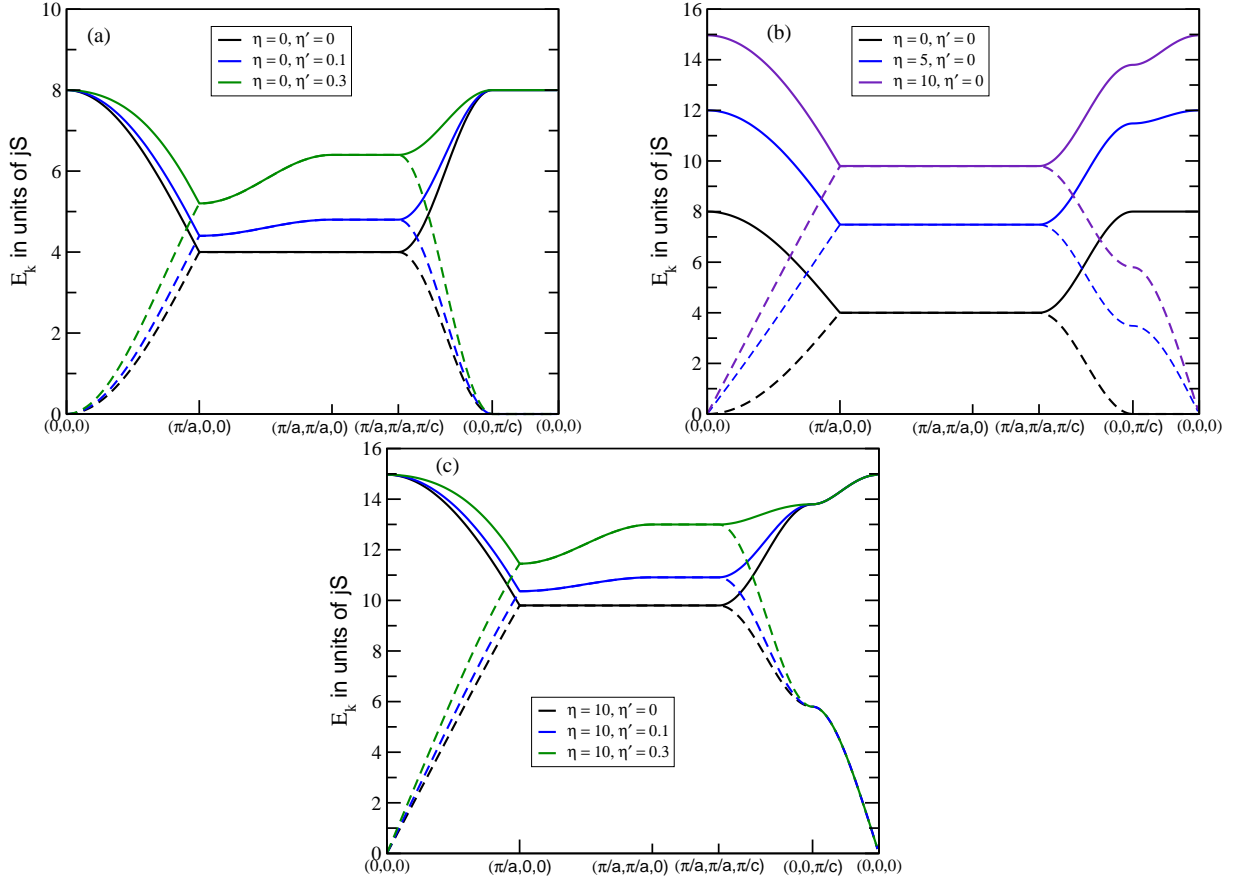

\centering
\includegraphics[width=3.0in,clip]{EnergyFAFa.eps}
\qquad 
\includegraphics[width=3.0in,clip]{EnergyFAFb.eps}
\qquad 
\qquad
\includegraphics[width=3.0in,clip]{EnergyFAFc.eps}
\caption{\label{fig:dispFAF} (Color online) (a) Magnon dispersion for acoustic (Goldstone) and optic magnons are shown for 
different values of intra-dimer coupling $\eta=J/j$
for NNN intra-bilayer ferromagnetic interaction $\eta^\prime=j^\prime/j=0$. (b, c) Effects of $\eta^\prime$ on the dispersion are 
for two different values of $\eta$. Each branch is two-fold degenerate corresponding to $\alpha$ and $\beta$ magnons.}
\end{figure}

Our tetragonal unit cell contains four Cr spins (two up and two down) -- so 
there are two $\alpha$ and two $\beta$ branches for each ${\bf k}$. 
Fig.~\ref{fig:dispFAF}a displays the magnon dispersions for $\eta=0$, 
when the two bilayers are decoupled. 
In the decoupled bilayer limit there are two magnon modes ($\omega _{\bf k}^{(1)}$ and $\omega _{\bf k}^{(2)}$) with different 
dispersions corresponding to the two spins per the two-dimensional square lattice unit cell $[a{\hat x},a{\hat y}]$. If we compare 
this dispersion 
with the case of Cr$_2$TeO$_6$ (Fig. 2 of Ref.~\onlinecite{majumdar18a}) where the bilayers are antiferromagnetic, we find
that the two modes are degenerate.  In order to understand this we look at a smaller $2D$ unit cell, a rotated square 
lattice $[(a/2)({\hat x +\hat y}),(a/2)(-{\hat x}+{\hat y})]$ for which the corresponding BZ is larger. The dispersion for the ferro (F) case in the smaller unit 
cell when mapped into the smaller BZ of the larger unit cell, gives the two modes seen in Fig.~\ref{fig:dispFAF}a. On the other 
hand because of the degeneracy between $\alpha$ and $\beta$ for the antiferro (AF) case, the mapping gives a two-fold degenerate 
mode. When we turn on 
the inter-bilayer AF coupling the $\alpha$ and $\beta$ degeneracy does not split the two modes into four modes. If on the 
other hand, the inter-bilayer couplings were ferromagnetic one would have seen four modes corresponding to four 
ferromagnetically oriented spins per unit cell.
The absence of $k_z$ dependence is obvious as with $\eta=0$ there is no coupling between the layers along the $z$-direction. Introduction of 
a nonzero NNN exchange 
coupling $\eta^\prime$ brings in dispersion along $(\pi/a,0,0)$ to $(\pi/a, \pi/a, 0)$.

In Fig.~\ref{fig:dispFAF}b we show the effect of 
introducing inter-bilayer 
AF coupling $\eta$ (for 
simplicity we chose $\eta^\prime=0$). Non-zero $\eta$ couples the intra-bilayer modes, leading to acoustic (Goldstone modes, 
$\omega_{\bf k}^{(1)} \rightarrow 0$ as ${\bf k}\rightarrow 0$) 
and optic modes ($\omega_{\bf k}^{(2)} \rightarrow 4/\sqrt{4+\eta}$ as ${\bf k}\rightarrow 0$). The new $\alpha$ and $\beta$ modes are linear 
combinations of the old decoupled bilayer  
modes. The modes split into two modes along $(0,0,0)$ to $(\pi/a,0,0)$ and the zero frequency modes along $(0,0,0)$ 
to $(0,0, \pi/c)$ split into acoustic and optic modes.  Interestingly, the modes along $(\pi/a,0,0)$ to $(\pi/a, \pi/a,0)$ to 
$(\pi/a, \pi/a, \pi/c)$ 
are dispersionless  and four-fold degenerate. Finally, in Fig.~\ref{fig:dispFAF}c, we show how the NNN ferromagnetic coupling 
introduces dispersion to these modes, but it does not remove the degeneracy.

It is interesting to compare the basic differences in the magnon dispersions for the AF-AF and F-AF cases for the same value 
of $\eta\; (=10)$. For simplicity we again consider the case $\eta^\prime=0$. Comparing Fig.~\ref{fig:dispFAF}c of the present paper with 
Fig.~2c of Ref.~\onlinecite{majumdar18a}, we see that there is a strong similarity between the dispersions from 
$(0,0,0) \rightarrow (\pi/a,0,0) \rightarrow ( \pi/a, \pi/a,0) \rightarrow ( \pi/a, \pi/a, \pi/c)$, with the exception of the width of 
the optical magnons along $(0,0,0) \rightarrow  (\pi/a,0,0)$. It is (in units of $jS$) about a factor of 2 larger for the F-AF case. 
The main difference is seen in the dispersions along $(0,0,0) \rightarrow (0,0, \pi/c) \rightarrow (\pi/a, \pi/a, \pi/c)$. Both 
the optic and acoustic branches are dramatically different.

As an example, consider the dispersions for the Goldstone mode for both F-AF and AF-AF (Ref.~\onlinecite{majumdar18a}) 
systems with non-zero small ${\bf k}$ (we kept $\eta^\prime=0$ for simplicity): 
\begin{subequations}
\bea
\omega_{\bf k}^{(1),{\rm F-AF}} &\approx& \Big[\frac{(k_x^2+k_y^2)a^2}{4(\eta+4)^2}\Big(4\eta+(k_x^2+k_y^2)a^2\Big)
+\frac {\eta^2}{(4+\eta)^3}(k_z c)^2 \Big]^{1/2},\label{GmodeFAF}\\
\omega_{\bf k}^{(1),{\rm AF-AF}} &\approx& \frac 1{\sqrt{4+\eta}}\Big[(k_x^2+k_y^2)a^2+\frac {\eta}{4+\eta} (k_zc)^2 \Big]^{1/2}.
\label{GmodeAFAF}
\eea
\end{subequations}
For small
$k_x, k_y$ with $k_z=0$ (dispersion in the basal plane) $\omega_{\bf k}^{(1),{\rm F-AF}} \rightarrow \frac {ka}{2(4+\eta)}\sqrt{4\eta+(ka)^2}$ where 
$k=(k_x^2+k_y^2)^{1/2}$. Clearly $\omega_{\bf k}^{(1),{\rm F-AF}} \rightarrow 0$ as $k\rightarrow 0$. For $ka << \sqrt{4\eta}$, the dispersion 
is linear corresponding to AF magnons which behave like ferromagnetic magnons for $ka >>\sqrt{4\eta}$. The crossover occurs for the
wave-vector 
$k_c \sim \sqrt{4\eta}/a$. As seen in Eq.~\eqref{GmodeFAF}  a quadratic dispersion for $\eta=0$ starts to
develop a linear term as $\eta$ becomes non-zero. 
We also note that for $k_x=k_y=0$ the dispersion is linear in $k_z$ with finite $\eta$ as seen in Fig.~\ref{fig:dispFAF} for the region 
$(0,0,\pi/c)\rightarrow (0,0,0)$. 
This is in sharp contrast to the AF-AF system (Eq.~\eqref{GmodeAFAF}) 
where a linear dispersion for $\eta=0$ remains linear when $\eta$ becomes finite (see Fig. 2b in Ref.~\onlinecite{majumdar18a}). 
Single crystal neutron scattering measurements should be able to detect these features.
\subsection{Sublattice Magnetization}
\begin{figure}[httb]
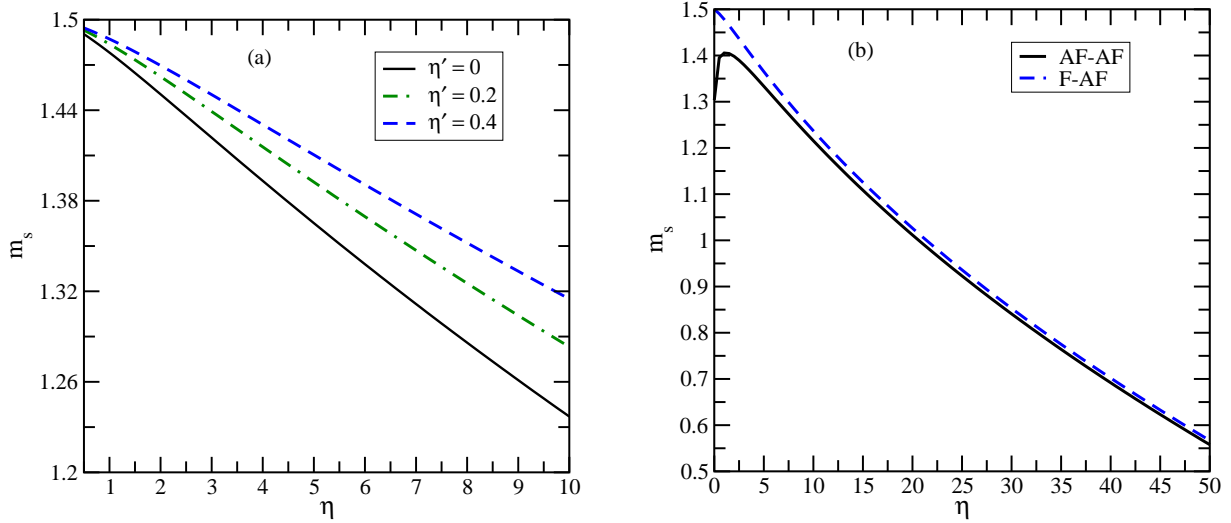

\centering
\includegraphics[width=3.0in,clip]{MagFAFa.eps}
\qquad 
\includegraphics[width=3.0in,clip]{MagFAFb.eps}
\caption{\label{fig:magFAF} (Color online) (a) Normalized sublattice magnetization, $m_s$ is shown as a function of the inter-bilayer coupling parameter 
$\eta$ for different values of NNN interaction $\eta^\prime$. 
(b) Comparison of $m_s$ as a function of the inter-bilayer coupling parameter 
$\eta$ is shown for AF-AF and F-AF bilayers.}
\end{figure}

We calculate the normalized sublattice magnetization $m_s=M_s/M_{0}$ from Eq.~\eqref{MagF} as a function of $\eta$. 
Fig.~\ref{fig:magFAF}a shows the magnetizations for F-AF bilayer as a function of $\eta$ and for different values of 
$\eta^\prime$. For $\eta^\prime=0$, magnetization starts from the classical value 1.5 (at $\eta=0$) and then monotonically decreases 
with increasing $\eta$. This is expected as increasing antiferromagnetic coupling $\eta$ enhances QSF and thus reduces 
$m_s$. However, adding ferromagnetic NNN interactions $\eta^\prime$ enhances $m_s$ -- this is shown in Fig.~\ref{fig:magFAF}a
for two different values of $\eta^\prime=0.2$ and $0.4$. 
On the other hand for AF-AF bilayer (as in Cr$_2$TeO$_6$ systems) $m_s$ increases from the initial value of 
1.303 (at $\eta=\eta^\prime=0$) to 1.406 (at $\eta=1.25$) and then decreases monotonically as shown in Fig.~\ref{fig:magFAF}b. 
Eventually for large value of $\eta$, $m_s$ for both AF-AF and F-AF bilayer approach each other.

\subsection{Two-Magnon Density of States (TM-DOS)}

\begin{figure}[httb]
\centering
\includegraphics[width=\textwidth,clip]{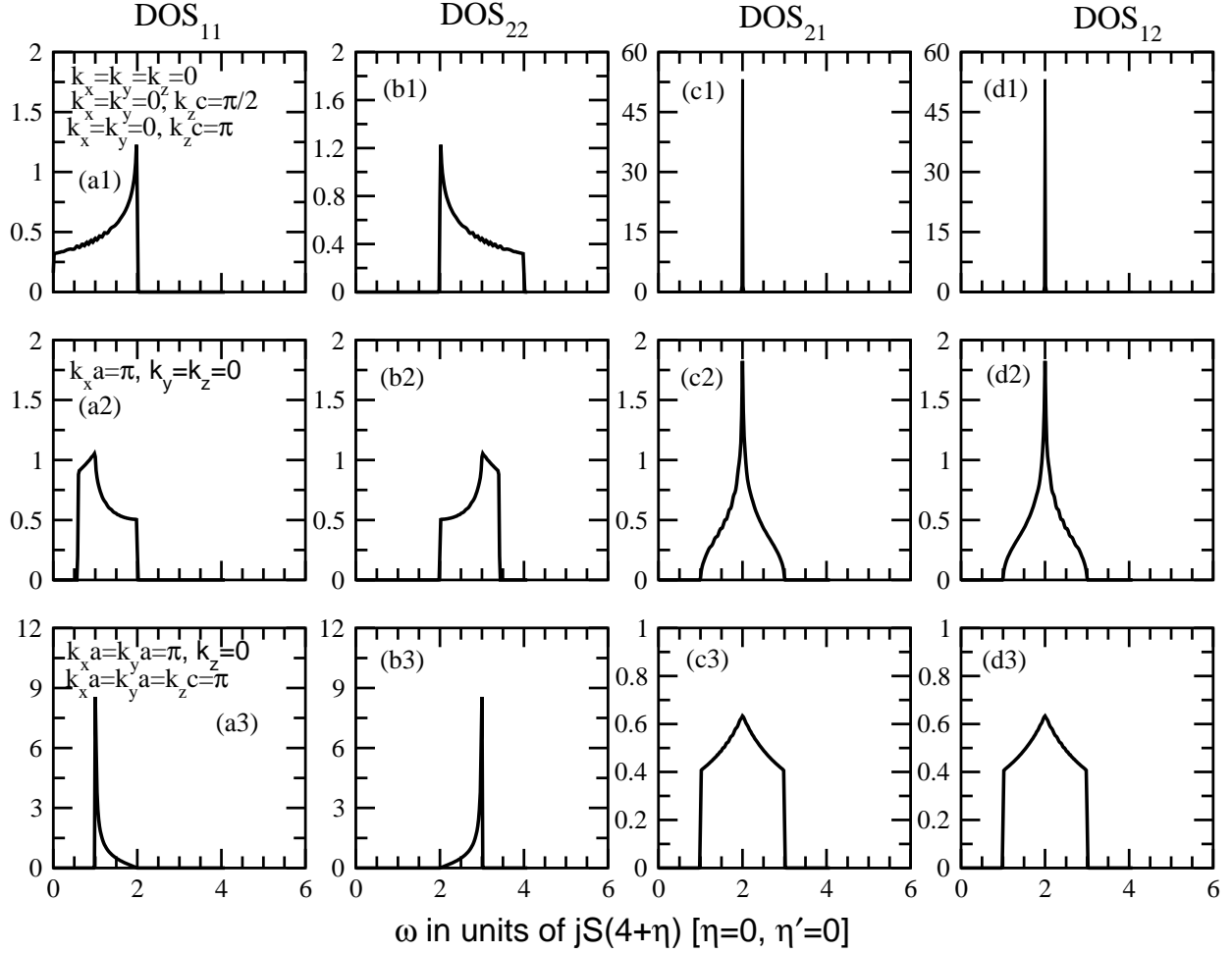}
\caption{\label{fig:FAFDOS1}(a1--d3) Two-magnon density of states for different values of ${\bf k}$ is plotted 
for $\eta=\eta^\prime=0$. Notice that DOS$_{12}({\bf k},\omega)=$DOS$_{21}({\bf k},\omega)$.}
\end{figure}
The longitudinal spin-spin correlation function ${\cal L}_s({\bf k},\omega)$, which is directly 
probed in 
inelastic scattering measurements depends sensitively on TM-DOS. The latter are 
calculated for different ${\bf k}$-values by numerically evaluating the internal 
three-dimensional momenta ${\bf p}$ on a mesh grid of size $L \times L \times L$, where $L = 256$. A Gaussian function 
of width $\sigma=0.075$ (in units of frequency $\omega$) is used to broaden the $\delta$-function. 
In Fig.~\ref{fig:FAFDOS1}, we present all four two-magnon DOS$({\bf k},\omega)$ for $\eta=0$. 
As discussed earlier, in the absence of inter-bilayer coupling one has ferromagnetic magnons associated with the two branches of 
the dispersion shown in Fig.~\ref{fig:dispFAF}a. Although the two intra-mode TM-DOS, DOS$_{11}({\bf k},\omega)$ and DOS$_{22}({\bf k},\omega)$ 
are different, the two inter-mode TM-DOS,  
DOS$_{21}({\bf k},\omega)$ and DOS$_{12}({\bf k},\omega)$ are equal.

\begin{figure}[httb]
\centering
\includegraphics[width=\textwidth,clip]{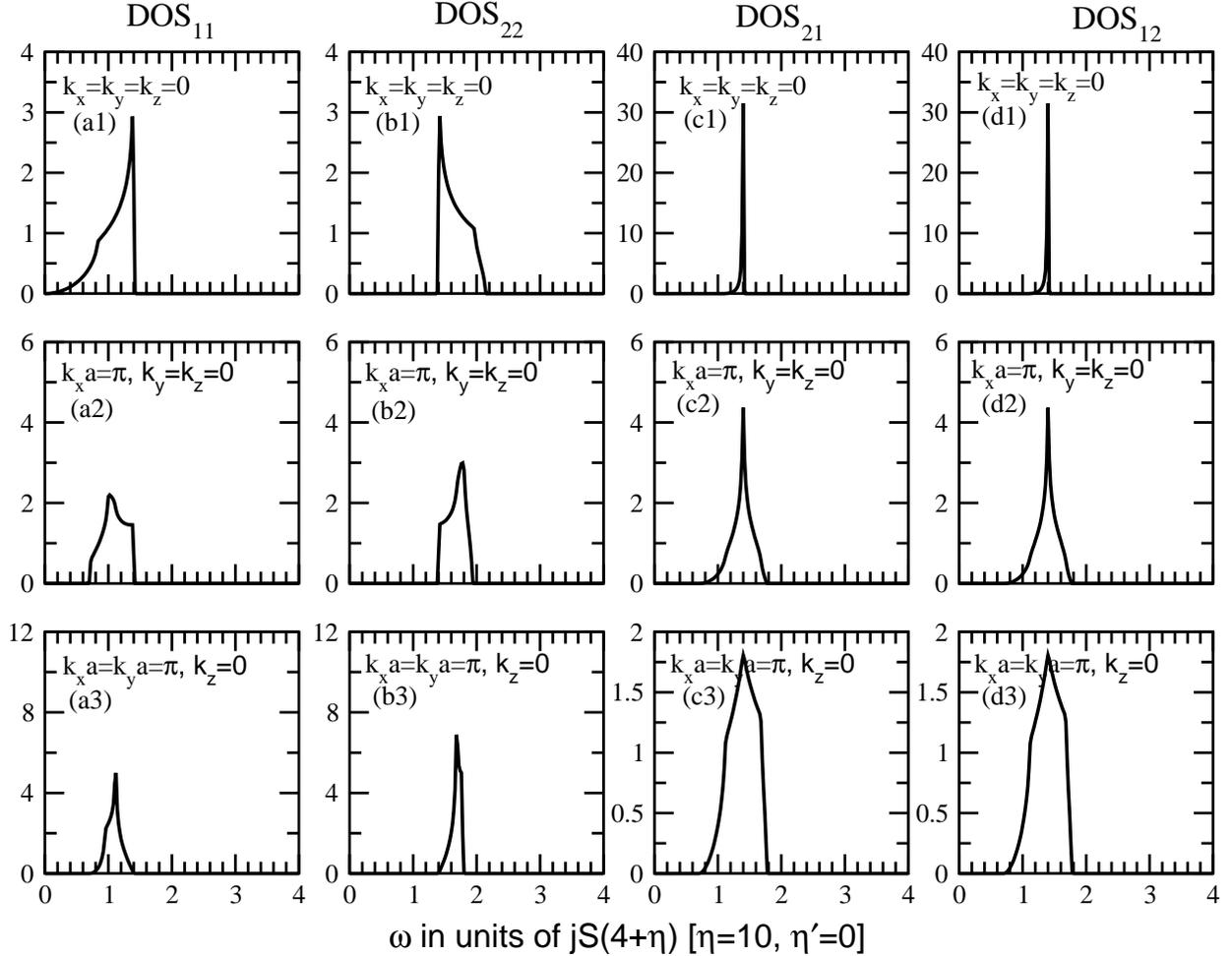}
\caption{\label{fig:FAFDOS2}(a1--d3) Two-magnon DOS for different values of $(k_x, k_y)$ with $k_z=0$ is plotted for $\eta=10, \eta^\prime=0$.
Notice that the symmetry DOS$_{12}({\bf k},\omega)=$DOS$_{21}({\bf k},\omega)$ still persists even for non-zero $\eta$.}
\end{figure}

\begin{figure}[httb]
\centering
\includegraphics[width=\textwidth,clip]{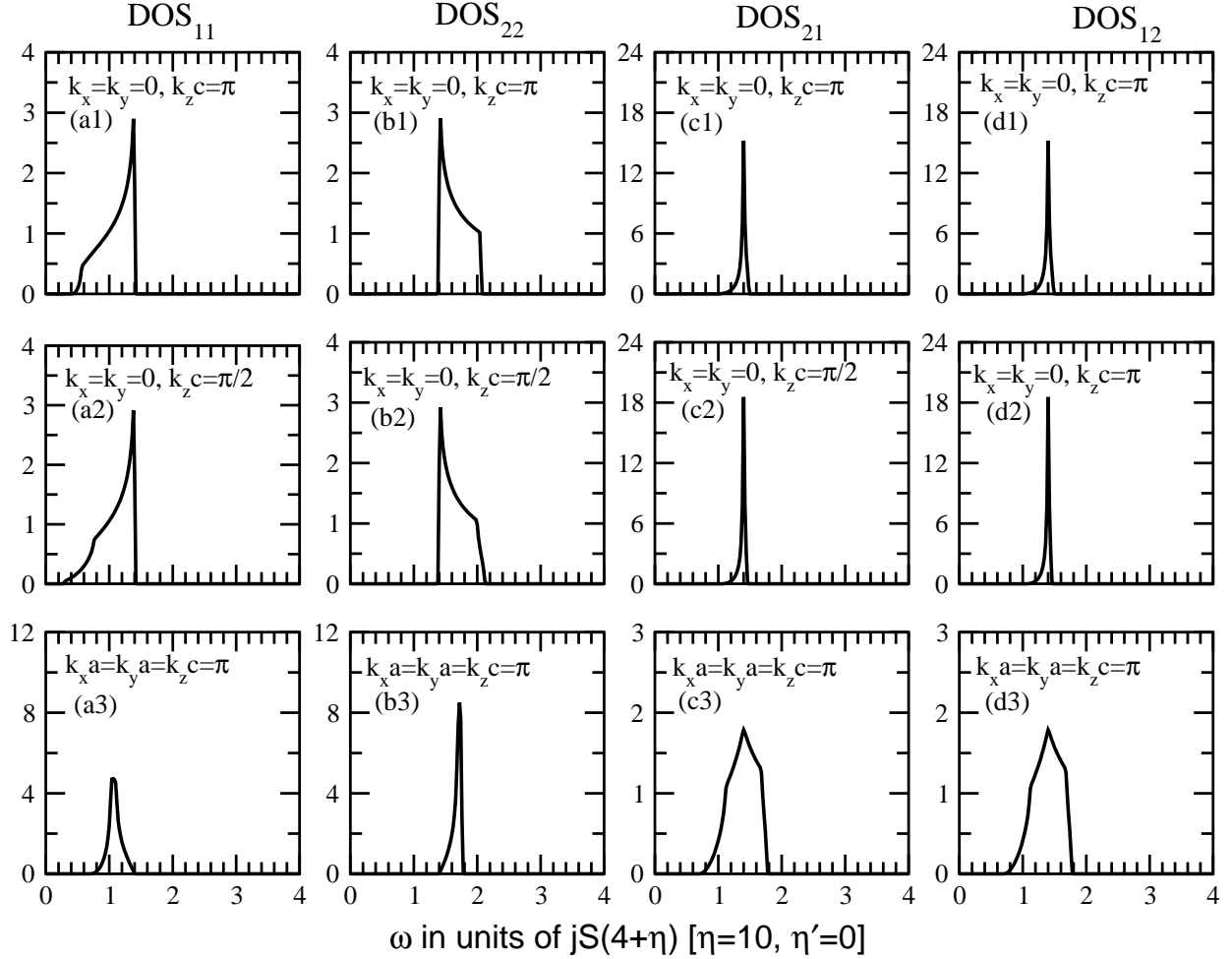}
\caption{\label{fig:FAFDOS3}(a1--d3) Two-magnon DOS for different values of $(k_x, k_y, k_z)$ is plotted for $\eta=10, \eta^\prime=0$. The plots
show the $k_z$ dependence on the four DOS.}
\end{figure}

Next, we discuss the case when inter-bilayer coupling is nonzero ($\eta \neq 0$). Since in the Cr$_2$(W, Mo)O$_6$ systems, $|J|$ is much larger 
than the intra-bilayer coupling $|j|$ we choose $\eta=10$ and still keep $\eta^\prime=0$ for simplicity. In 
Fig.~\ref{fig:FAFDOS2} and Fig.~\ref{fig:FAFDOS3}, we plot the 
$({\bf k},\omega)$ dependence of DOS$_{11}$, DOS$_{22}$, DOS$_{21}$, and DOS$_{12}$. The equality 
${\rm DOS}_{12}({\bf k},\omega) ={\rm DOS}_{21}({\bf k}, \omega)$ for any ${\bf k}$ is still preserved for non-zero $\eta$. 

In Fig.~\ref{fig:FAFDOS2}, we choose $k_z=0$ and study the $(k_x,k_y)$ dependence and in Fig.~\ref{fig:FAFDOS3}, we show the effect 
of $k_z$ on all four TM-DOS. Consider the evolution of the four TM-DOS as a function of $k_z$ with
$k_x=k_y=0$ as shown in Fig.~\ref{fig:FAFDOS2}a1-d1, Fig.~\ref{fig:FAFDOS3}a1-d1, and Fig.~\ref{fig:FAFDOS3}a2-d2. 
Especially consider the peak intensity (at 19.6$jS$) for the inter-band density of states DOS$_{12}$ (=DOS$_{21}$). The intensity 
is $\sim 32$ for
$k_x=k_y=k_z=0$ [Fig.~\ref{fig:FAFDOS2}c1] whereas it decreases to $\sim 15$ (at 19.6$jS$) for $k_x=k_y=0,k_z=\pi$ 
[Fig.~\ref{fig:FAFDOS3}c1]. Interestingly there is no change in the peak intensity for DOS$_{11}$ and DOS$_{22}$
[Fig.~\ref{fig:FAFDOS3}a1-b1].

\subsection{Longitudinal spin-spin correlation function (LSSCF)}
\begin{figure}[httb]
\centering
\includegraphics[width=\textwidth,clip]{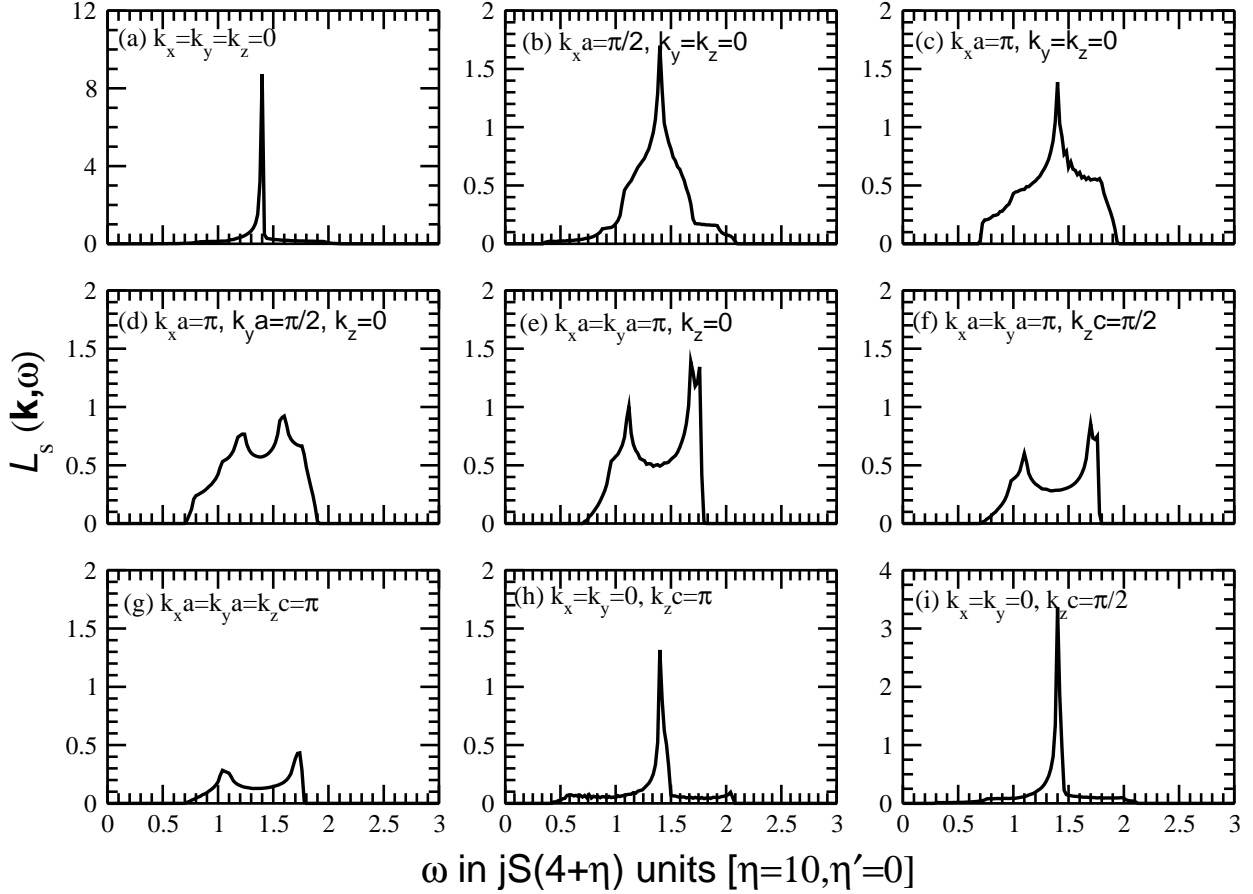}
\caption{\label{fig:LsFAF}(a--i) Longitudinal spin-spin correlation, ${\cal L}_{s}({\bf k}, \omega)$ for different values of ${\bf k}$ 
is plotted for $\eta=10, \eta^\prime=0$.}
\end{figure}
\begin{figure}[httb]
\centering
\includegraphics[width=\textwidth,clip]{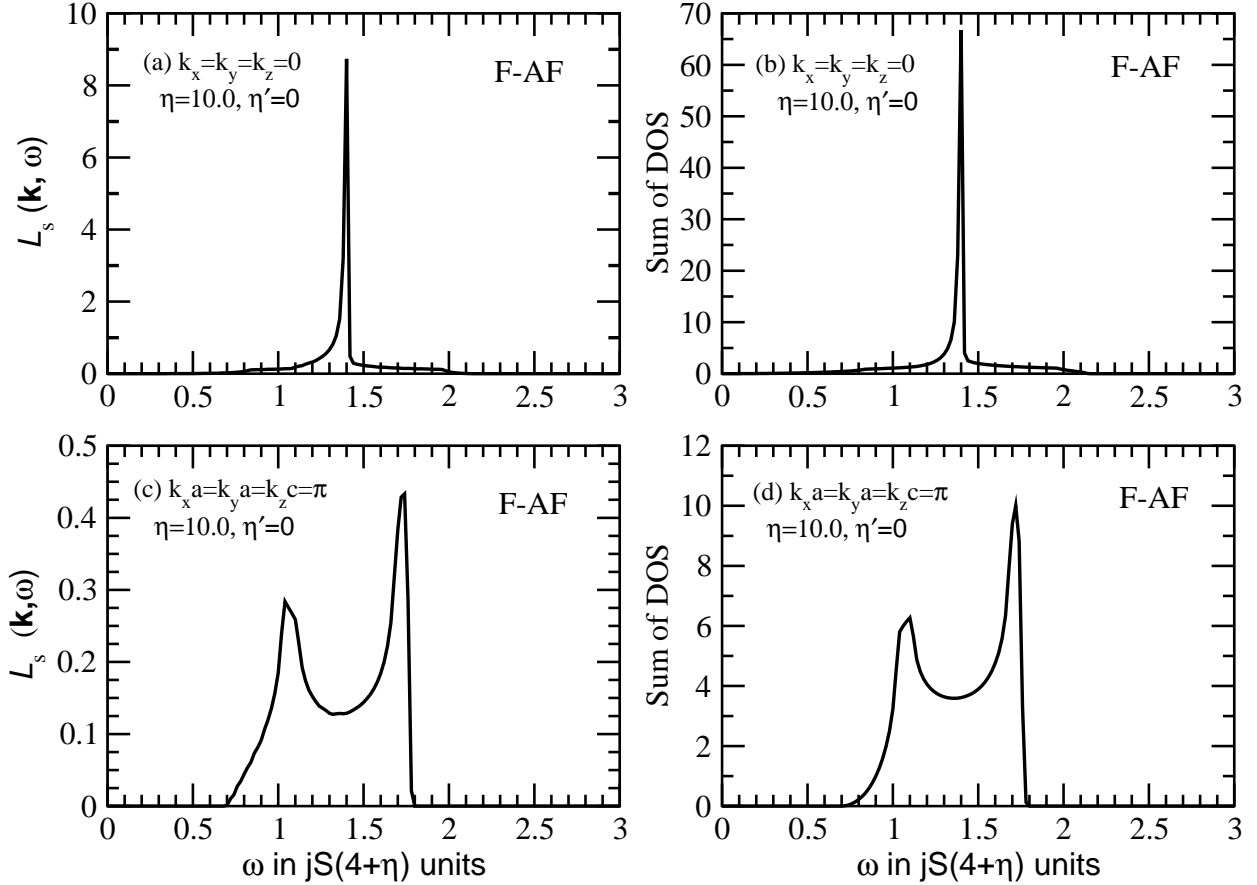}
\caption{\label{fig:SzzDOS} (a -- d) Longitudinal spin-spin correlation ${\cal L}_{s}({\bf k}, \omega)$ and the sum of the density of 
states of 
four magnon branches are plotted for $\eta=10.0, \eta^\prime=0$ or two different values of ${\bf k}$. The plots show 
the effect of the weights ${\cal D}_{ij}$ in ${\cal L}_{s}({\bf k}, \omega)$.}
\end{figure}

In Fig.~\ref{fig:LsFAF}a-i, we show the ${\bf k}$-dependence of LSSCF ${\cal L}_s({\bf k},\omega)$. As seen in 
Eq.~\eqref{spin-corr}, contributions from different two-magnon 
excitations get weighted by the associated weights ${\cal D}^{ij}_{{\bf k},{\bf k+p}}$. This leads to different energy dependence of 
LSSCF compared to that of the total two-magnon DOS.  In Fig.~\ref{fig:SzzDOS} we show both ${\cal L}_s({\bf k},\omega)$ and the sum of 
the four DOS$_{ij}({\bf k}, \omega)$ 
for ${\bf k}=0$ and $k_xa=k_ya=k_zc=\pi$. Both ${\cal L}_s({\bf k},\omega)$ and $\sum$DOS$_{ij}({\bf k}, \omega)$ show similar
features. However the intensity of the peak in ${\cal L}_s({\bf k},\omega)$ is reduced significantly, which shows
the effect of the weights. Another interesting feature is that ${\cal L}_s({\bf k},\omega)$ has only one peak at 
$\omega=1.4jS(4+\eta)=19.5jS$
for ${\bf k}=0$ [Fig.~\ref{fig:LsFAF}a], whereas for $k_xa=k_ya=k_zc=\pi$ two peaks emerge, one at $\omega=1.0jS(4+\eta)=14.6jS$ and 
the other at $\omega=1.74jS(4+\eta)=24.4jS$ [Fig.~\ref{fig:LsFAF}g]. The formation of two peaks from a single peak in ${\cal L}_s({\bf k},\omega)$ can be 
seen in Fig.~\ref{fig:LsFAF}. 

\begin{figure}[httb]
\centering
\includegraphics[width=\textwidth,clip]{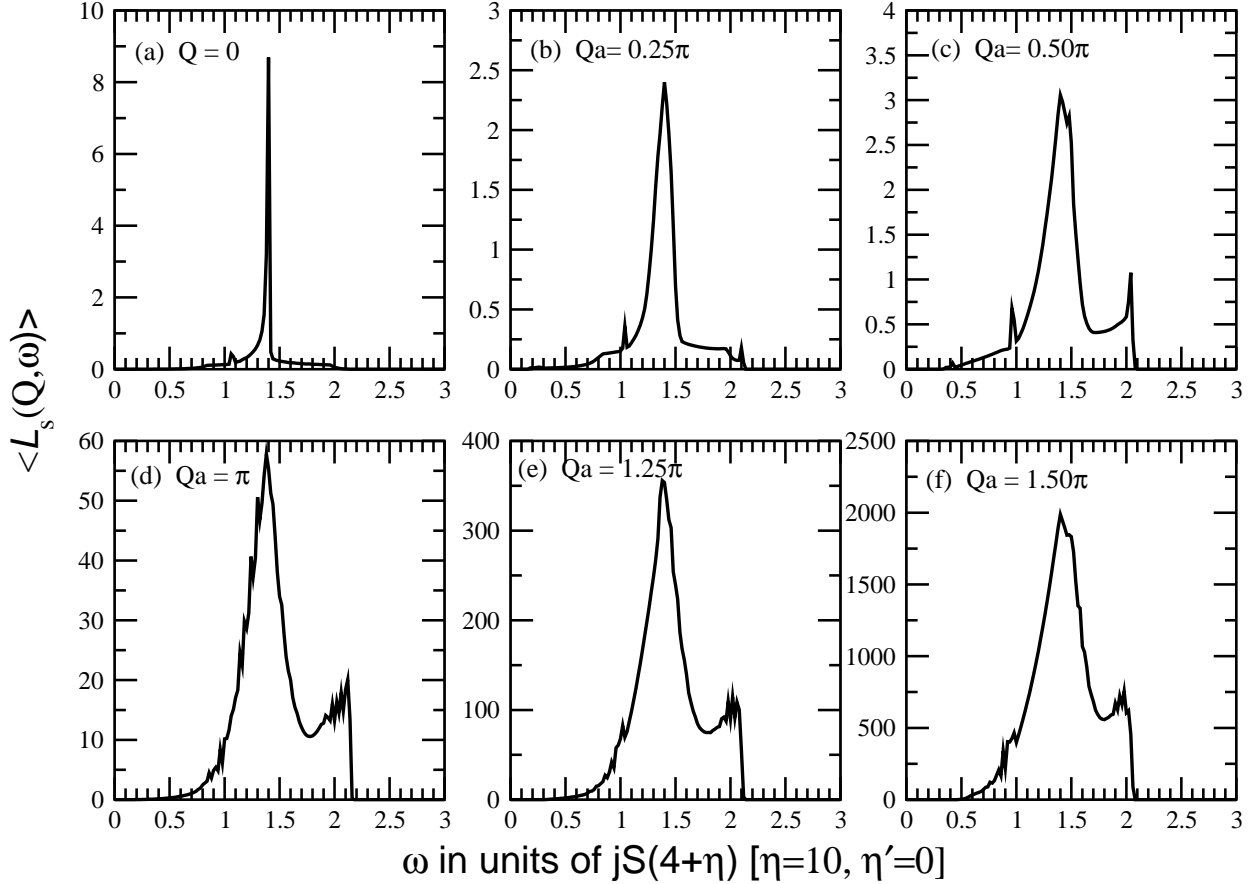}
\caption{\label{fig:PowAvg} (a -- f) Powder-averaged longitudinal spin-spin correlation 
function $\langle {\cal L}_s(|{\bf k}|=Q,\omega)\rangle$ for $\eta=10, \eta^\prime=0$ and 
$Qa=0, 0.25\pi, 0.5\pi, \pi$. }
\end{figure}

Finally, we plot the angular average of ${\cal L}_s({\bf k},\omega)$ for different magnitudes of $|{\bf k}|=Q$ in Fig.~\ref{fig:PowAvg}. 
For these plots, Eq.~\eqref{powavg} was numerically 
evaluated by summing over the angles $\theta,\phi$. 
For each $\omega$ about 270 million points were evaluated. This is what can be observed in a inelastic neutron scattering experiment
from a powder sample. The generic feature is a narrow peak seen at $\sim 19.5jS$ with two broad peaks on 
each side.
With increase in the magnitude of ${\bf k}$ the intensity initially decreases (from $Qa=0$ to $Qa=0.50\pi$) and then 
increases (from $Qa=\pi$ to
$Qa=1.5\pi$). Moreover the broad peak at $\omega=2.0jS(4+\eta)= 28jS$ increases in intensity with increase in $Q$.

\section{\label{sec:compare}Comparison between F-AF and AF-AF bilayer systems}
In our previous work we studied the low-temperature magnetic properties of the Cr$_2$TeO$_6$ bilayer system 
where both the intra and inter-bilayer couplings are antiferromagnetic.~\cite{majumdar18a}
In this paper we have discussed the magnon dispersion, two-magnon density of states, and longitudinal spin-spin correlation 
function in the leading order approximation for Cr$_2$WO$_6$ 
and Cr$_2$MoO$_6$ coupled bilayer systems
where inter-bilayer NN coupling is antiferromagnetic but intra-bilayer coupling is ferromagnetic. We have also investigated 
how a small intra-bilayer NNN ferromagnetic coupling affects the above properties.

We find that F-AF system differs in several ways from the AF-AF system studied in the earlier paper.~\cite{majumdar18a} 
\begin{enumerate}
\item For the  F-AF bilayer 
system the two magnon branches with frequencies $\omega_{\bf k}^{(1)}$ and  
$\omega_{\bf k}^{(2)}$ corresponding  
to two bilayers for $\eta=0$ are non-degenerate [Fig.~\ref{fig:dispFAF}] except between $(\pi/a,0,0)$ to
$(\pi/a, \pi/a,0)$ to $(\pi/a, \pi/a, \pi/c)$ (within LSWT - higher order $1/S$ corrections may lift this degeneracy). This result 
is different from the AF-AF case (for Cr$_2$TeO$_6$ systems) where both the branches are degenerate throughout the first BZ [Fig. 2 
of Ref.~\onlinecite{majumdar18a}]. 
Also as we pointed out earlier, for large values of $\eta (=10)$, the magnon dispersions are very similar 
along $(0,0,0) \rightarrow (\pi/a,0,0) \rightarrow (\pi/a,\pi/a,0) \rightarrow (\pi/a,\pi/a,\pi/c)$ for the two cases, but 
differ dramatically from $(0,0,0) \rightarrow (0,0,\pi/c) \rightarrow  (\pi/a,\pi/a,\pi/c)$.

As another example, the dispersions for the Goldstone mode between F-AF and AF-AF systems are quite different as seen 
in Eqs.~\eqref{GmodeFAF}-\eqref{GmodeAFAF}. For the F-AF system a quadratic dispersion for $\eta=0$ starts to
develop a linear term as $\eta$ becomes non-zero (see Eq.~\eqref{GmodeFAF}) whereas for the AF-AF system
a linear dispersion for $\eta=0$ remains linear when $\eta$ becomes finite (see Eq.~\eqref{GmodeAFAF}).

\item The normalized sublattice magnetization $m_s$ for both F-AF and AF-AF bilayer systems differs substantially from 
its classical value due to quantum spin fluctuations with increase in $\eta$ [Fig.~\ref{fig:magFAF}b]. 
In case of F-AF bilayers $m_s$ start from the classical value of 1.5 (at $\eta=0$) and 
then monotonically decreases with increasing $\eta$.  
On the contrary, for the AF-AF bilayer system, we have found a non-monotonic $\eta$ dependence of $m_s$ -- it initially 
increases from the initial value of 1.303 at $\eta=\eta^\prime=0$  
to 1.406 (at $\eta=1.25$) and then decreases monotonically. Eventually for large values 
of $\eta$, $m_s$ for both AF-AF and F-AF bilayers become identical.  
Addition of ferromagnetic NNN interaction $j^\prime$ suppresses QSF effects and thereby enhances $m_s$ in both cases.

\begin{figure}[httb]
\centering
\includegraphics[width=\textwidth,clip]{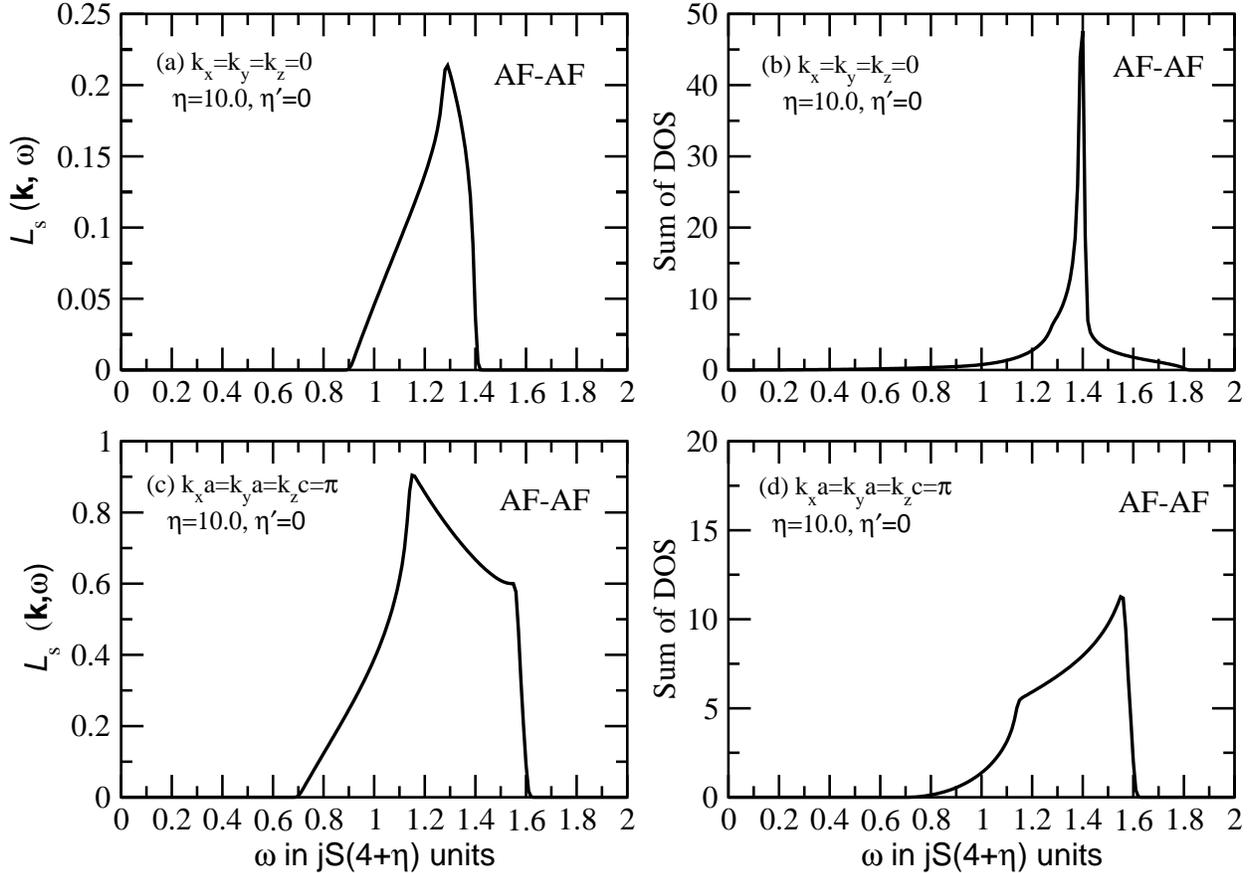}
\caption{\label{fig:SzzDOSAF} Longitudinal spin-spin correlation ${\cal L}_{s}({\bf k}, \omega)$ and the sum of the density of states of 
four magnon branches are plotted for $\eta=10.0, \eta^\prime=0$ or two different values of ${\bf k}$ for the AF-AF system. The plots show 
the effect of the weights ${\cal D}_{ij}$ in ${\cal L}_{s}({\bf k}, \omega)$.}
\end{figure}

\item There are some differences for the two-magnon DOS between AF-AF and F-AF bilayer systems with 
$\eta=10, \eta^\prime=0$ [Fig.~\ref{fig:FAFDOS2}-\ref{fig:FAFDOS3} and Figs. 5 and 6 in Ref.~\onlinecite{majumdar18a}]. 
As an example, 
for ${\bf k}=0$, DOS$_{12}$=DOS$_{21}$ for both the systems. But for $k_x=k_y=0, k_zc=\pi$, two inter-band DOS are equal
to their corresponding two intra-band DOS i.e. DOS$_{11}$=DOS$_{12}$ and 
DOS$_{22}$=DOS$_{21}$ for AF-AF bilayers whereas for F-AF bilayers only the two inter-band DOS are equal, i.e.  
DOS$_{12}$=DOS$_{21}$. Another interesting observation is that for the AF-AF bilayers ${\cal L}_s({\bf k},\omega)=0$ at 
$k_xa=k_ya=\pi,k_z=0$ even though all the density of states are non-zero [Fig. 7e in Ref.~\onlinecite{majumdar18a}]. But with the 
F-AF bilayers, we have not found any ${\bf k}$ for which ${\cal L}_s({\bf k},\omega)$ vanishes with non-zero DOS. 

\item Comparison of the sum of the density of states (right panel) with LSSCF (left panel) in 
Figs.~\ref{fig:SzzDOS} and \ref{fig:SzzDOSAF} show the effect of the wave-functions on  ${\cal L}_{s}({\bf k}, \omega)$. 
We observe from Fig.~\ref{fig:SzzDOSAF} that for the AF-AF system 
the wave-functions 
substantially changes the LSSCF structure from the sum of DOS. On the contrary, for the F-AF system in Fig.~\ref{fig:SzzDOS} the change
in the structure of LSSCF from the sum of DOS is minimal (other than an overall reduction in the peak intensity).

\item Finally for the 
powder average we find only a narrow peak seen at $19.5jS$ [at $\omega=1.4jS(4+\eta)$] with a small broad peak at 
lower energies for the AF-AF system [Fig. 9 of Ref.~\onlinecite{majumdar18a}]. But, for the F-AF system a narrow peak is seen 
at $\sim 19.5jS$ 
[at $\omega=1.4jS(4+\eta)$] with two broad peaks on each side [Fig.~\ref{fig:PowAvg}(c--f)]. However the broad peak 
at $\sim 28jS$ [at $\omega=2jS(4+\eta)$] 
increases in intensity with increase in $|k|=Q$. One more difference is that for the AF-AF system the intensity increases with 
increase in the magnitude of ${\bf k}$, whereas for the F-AF system it first decreases and then increases as we approach 
the zone boundary at $Q=1.5\pi/a$.
\end{enumerate}

\section{\label{sec:conclusions}Conclusions}

In this article  we studied the magnetic properties (magnon dispersion, suppression of long range order by quantum spin fluctuation, 
two-magnon density of states, longitudinal spin-spin correlation function and its angular average) of Cr$_2$WO$_6$ and Cr$_2$MO$_6$,
which are bilayer systems of antiferromagnetically coupled (strength $J$) quantum spin-3/2 dimers interacting through 
$2D$ ferromagnetic coupling (strength $j$). In addition to $J$ and $j$, there is also a small 
inter-dimer longer range ferromagnetic coupling ($j^\prime$) whose magnitude is much smaller than $J$ and $j$. For convenience
we will consider $j^\prime=0$.
In a recent paper [Ref.~\onlinecite{majumdar18a}], we discussed 
the magnetic properties of a related system, Cr$_2$TeO$_6$, where the dimers are coupled antiferromagnetically. 
There are many similarities and differences between the two cases (F-AF and AF-AF).  
In the limit $J=0$, W and Mo systems reduce to non-interacting $2D$ ferromagnetic (F) sheets whereas the Te system 
reduces to non-interacting $2D$ antiferromagnetic (AF) sheets. The magnon dispersions are therefore qualitatively different, 
for small ${\bf k}$ (linear for AF and quadratic for F sheets). In addition the total magnon band-width (in unit of $j$S) is 4 
for AF and 8 for F. However, the intra-dimer AF coupling is dominant, $J\sim 10j$, and it controls the magnon 
dispersion. In this limit, magnon dispersions are qualitatively similar excepting along the directions $(0,0,0)$ to $(0,0,\pi/c)$ 
and along $(0,0, \pi/c)$ to $(\pi/a, \pi/a, \pi/c)$ (see Fig. 2(c) of Ref.~\onlinecite{majumdar18a} and Fig.~\ref{fig:dispFAF}c of 
this paper, for $\eta^\prime=0$). In the case of 
intra-layer AF coupling, the inter-layer AF coupling introduces two magnon modes (acoustic and optic) propagating along 
the $c$-axis which become degenerate at $(0,0,\pi/c)$. In contrast, for intra-layer F coupling, there is a large gap between 
the acoustic and optic modes $(\sim 7jS)$ at $(0,0,\pi/c)$. Careful single-crystal inelastic neutron scattering measurements should 
be able to detect these subtle differences between the Te system and W/Mo system. Quantum spin fluctuations (QSF) 
suppress the ordered maagnetization $(M_s/M_0)$ from its classical value $3/2$ for both F-AF and AF-AF systems.
In W/Mo systems, $(M_s/M_0)$ reduces monotonically from 
the classical value as $\eta$ increases. On the other hand, for the Te case, QSF already reduce $(M_s/M_0)$ when $\eta=0$. 
Introduction of $\eta$ first 
suppresses QSF and enhances the magnetization and then for larger $\eta$ values it decreases monotonically similar to the 
F-AF system. In case of F-AF system for $\eta=10$, $M_s/M_0 \sim 1.23-1.25$, which is 
about 17-18\% reduction (see Fig.~\ref{fig:magFAF}b). Finally for the angle averaged longitudinal spin-spin correlation function
$\langle {\cal L}_{s}(Q, \omega)\rangle $ the scattering intensity 
is a factor of 10 stronger for the W/Mo system compared to the Te system, again a result which can be verified experimentally.

\section{Acknowledgment}
We acknowledge the use of HPC cluster at GVSU, supported by the National Science Foundation 
Grant No. CNS-1228291 that have contributed to the research results reported within this paper. SDM would like 
to thank Dr. Xianglin Ke for stimulating discussions.

\appendix

\section{\label{formalism} Brief derivation of the Hamiltonian in momentum space}
The spin Hamiltonian in Eq.~\eqref{ham} is mapped onto a Hamiltonian of interacting bosons by expressing the spin operators in terms of 
bosonic creation 
and annihilation operators $a^\dag, a$ for ``up'' sites on sublattice A (and $b^\dag, b$ for ``down'' sites on  sublattice 
B)  using the  Holstein-Primakoff representation~\cite{HP}
\begin{subequations}\label{holstein}
\begin{eqnarray}
S_{in}^{+ A} &\approx& \sqrt{2S}a_{in},\;\;\;
S_{in}^{- A} \approx \sqrt{2S}a_{in}^\dag,\;\;\;
S_{in}^{z A} = S-a^\dag_{in} a_{in}, \label{hol1} \\ 
S_{jn}^{+ B} &\approx& \sqrt{2S}b_{jn}^\dag,\;\;\;
S_{jn}^{- B} \approx \sqrt{2S}b_{jn}, \;\;\;
S_{jn}^{z B} = -S+b^\dag_{jn}b_{jn}. \label {hol2}
\end{eqnarray}
\end{subequations}
After substituting Eqs.~\eqref{holstein} into Eq.~\eqref{ham} and 
expanding the Hamiltonian perturbatively in powers of $1/S$ (up to the quadratic term) we obtain:
\be
{\cal H}= {\cal H}_{\rm cl}+{\cal H}_0 + \cdots,
\ee 
where,
\begin{subequations}
 \label{HFF} 
\begin{eqnarray}
{\cal H}_{\rm cl} &=& -2jNS^2(4+\eta),\label{HF1} \\
{\cal H}_{0} &=& jS\sum_{n=1}^{N_z} \sum_{\langle i,j\rangle} \Big[a_{in}^{(1)\dag}a_{in}^{(1)}+a_{in}^{(2)\dag}a_{in}^{(2)} 
+b_{jn}^{(3)\dag}b_{jn}^{(3)}+b_{jn}^{(4)\dag}b_{jn}^{(4)} \non \\
&-& a_{in}^{(1)}a_{jn}^{(2)\dag}-a_{in}^{(1)\dag}a_{jn}^{(2)}
-b_{in}^{(3)\dag}b_{jn}^{(4)}-b_{in}^{(3)}b_{jn}^{(4)\dag}\Big] \non \\
&+& JS\sum_{n=1}^{N_z} \sum_{i} \Big[a_{in}^{(1)\dag}a_{in}^{(1)}+b_{in}^{(3)\dag}b_{in}^{(3)} 
+ a_{in}^{(1)}b_{in}^{(3)}+a_{in}^{(1)\dag}b_{in}^{(3)\dag} \non \\
&+& \frac 1{2}\Big\{ a_{in}^{(2)\dag}a_{in}^{(2)}+a_{in+1}^{(2)\dag}a_{in+1}^{(2)}
+b_{in}^{(4)\dag}b_{in}^{(4)}+b_{in-1}^{(4)\dag}b_{in-1}^{(4)}  \non \\
&+& a_{in}^{(2)}b_{in-1}^{(4)}+a_{in}^{(2)\dag}b_{in-1}^{(4)\dag} 
+a_{in+1}^{(2)}b_{in}^{(4)}+a_{in+1}^{(2)\dag}b_{in}^{(4)\dag}\Big\}\Big]\non \\
&+&j^\prime S\sum_{n=1}^{N_z} \sum_{\langle\langle i,j\rangle\rangle} \sum_{p=1,2}\Big[a_{in}^{(p)\dag}a_{in}^{(p)}+a_{jn}^{(p)\dag}a_{jn}^{(p)} 
-a_{in}^{(p)\dag}a_{jn}^{(p)}-a_{in}^{(p)}a_{jn}^{(p)\dag}\Big] \non \\
&+& j^\prime S\sum_{n=1}^{N_z} \sum_{\langle\langle i,j\rangle\rangle} \sum_{p=3,4}\Big[b_{in}^{(p)\dag}b_{in}^{(p)}+b_{jn}^{(p)\dag}b_{jn}^{(p)} 
-b_{in}^{(p)\dag}b_{jn}^{(p)}-b_{in}^{(p)}b_{jn}^{(p)\dag}\Big].\label{HF2}
\end{eqnarray}
\end{subequations}
${\cal H}_{\rm cl}$
represents the classical ground state (mean-field) energy and it is not relevant for the quantum fluctuations,
so we do not discuss it further. $H_0$ in Eq.~\eqref{HF2} is the quadratic part of the Hamiltonian.
In Eq.~\eqref{HF1}, the parameters $\eta=J/j$, $\eta^\prime=j^\prime/j$ and $N=N_xN_yN_z$ is the total number of unit cells. Next the 
real space Hamiltonian 
is transformed to momentum space using the Fourier transformation for each $\ell$-th spin:
\be
a_{in}^{(\ell)}=\frac 1{\sqrt{N}}\sum_{{\bf k}}e^{i{\bf k \cdot R_{in}^{(\ell)}}}a_{\bf k}^{(\ell)},\;\;\;
b_{in}^{(\ell)}=\frac 1{\sqrt{N}}\sum_{{\bf k}}e^{-i{\bf k \cdot R_{in}^{(\ell)}}}b_{-\bf k}^{(\ell)}.
\ee
Furthermore we have rescaled the operators $a,\;b$ as 
\bea
a\k^{(1)} &\equiv& e^{-ik_z \delta/2} a\k^{(1)},\;\;
a\k^{(4)}\equiv e^{-ik_z \delta/2} a\k^{(4)}, \non \\
b\km^{(2)}&\equiv& e^{-ik_z \delta/2} b\km^{(2)},\;\;
b\km^{(3)}\equiv e^{-ik_z \delta/2} b\km^{(3)}, \non 
\eea
where $\delta$ is the inter-dimer separation (Fig.~\ref{fig:CrMWstruc1}). In momentum space the quadratic Hamiltonian 
is shown in Eq.~\eqref{quadH}. 

\section{\label{lmcoeffs} Coefficients for Bogoliubov transformation}
First we define the following functions:
\begin{subequations}
\bea
U_{1\bf k} &=& -(1+\omega_{\bf k}^{(1)})(1+\vert \gamma_{1\bf k} \vert^2-\gamma_{2\bf k}^2-\omega^{(1)2}_{\bf k})
+2 \vert \gamma_{1\bf k} \vert^2,  \\
U_{1\bf k}^\prime &=& -(1+\omega_{\bf k}^{(2)})(1+\vert \gamma_{1\bf k} \vert^2 -\gamma_{2\bf k}^2 -\omega^{(2) 2}_{\bf k})
+2\vert \gamma_{1\bf k} \vert^2,  \\
U_{2\bf k} &=& -\gamma_{1\bf k}[(1+\omega_{\bf k}^{(1)})^{2}-\vert \gamma_{1\bf k} \vert^2]-\gamma_{1\bf k}^* \gamma_{2\bf k}^2,  \\
U_{2\bf k}^\prime &=& -\gamma_{1\bf k}[(1+\omega_{\bf k}^{(2)})^2-\vert \gamma_{1\bf k} \vert^2]-\gamma_{1\bf k}^* \gamma_{2\bf k}^2,  \\
V_{1\bf k} &=& (\gamma_{1\bf k}+\gamma_{1\bf k}^*) \gamma_{2\bf k}+\omega_{\bf k}^{(1)}
(\gamma_{1\bf k}-\gamma_{1\bf k}^*) \gamma_{2\bf k},  \\
V_{1\bf k}^\prime &=& (\gamma_{1\bf k}+\gamma_{1\bf k}^*) \gamma_{2\bf k})+\omega_{\bf k}^{(2)}
(\gamma_{1\bf k}-\gamma_{1\bf k}^*) \gamma_{2\bf k},  \\
V_{2\bf k} &=& \gamma_{2\bf k} (1-\gamma_{2\bf k}^2-\omega^{(1) 2}_{\bf k})+\gamma_{1\bf k}^{2}\gamma_{2\bf k},  \\
V_{2\bf k}^\prime &=& \gamma_{2\bf k} (1-\gamma_{2\bf k}^2-\omega^{(2) 2}_{\bf k})
+\gamma_{1\bf k}^{2}\gamma_{2\bf k},
\eea
\end{subequations}
then the coefficients for the BG transformaitons are --
\begin{subequations}
\bea
\ell_{1\bf k} &=& U_{1\bf k}/N_{1\bf k},\;\;
\ell_{1\bf k}^\prime = U_{1\bf k}^\prime/N_{2 \bf k},\;\;
\ell_{2\bf k} = U_{2\bf k}/N_{1\bf k},\;\;
\ell_{2\bf k}^\prime =  U_{2\bf k}^\prime/N_{2 \bf k},  \\
m_{1\bf k} &=& V_{1\bf k}/N_{1\bf k},\;\;
m_{1\bf k}^\prime = V_{1\bf k}^\prime/N_{2 \bf k}, \;\;
m_{2\bf k} = V_{2\bf k}/N_{1\bf k}, \;\;
m_{2\bf k}^\prime = V_{2\bf k}^\prime /N_{2 \bf k},
\eea
\end{subequations}
where the normalization factors $N_{1\bf k}, N_{2 \bf k}$ are given by:
\begin{subequations}
\bea
N_{1\bf k}&=& \Big[\vert U_{1 \bf k}\vert^2- \vert V_{1 \bf k}\vert^2
+\vert U_{2 \bf k}\vert^2-\vert V_{2 \bf k}\vert^2\Big]^{1/2},  \\
N_{2\bf k}&=& \Big[\vert U_{1 \bf k}^\prime\vert^2- \vert V_{1 \bf k}^\prime\vert^2
+\vert U_{2 \bf k}^\prime \vert^2-\vert V_{2 \bf k}^\prime\vert^2\Big]^{1/2}.
\eea
\end{subequations}

\section{\label{SScorr} Total spin $S_z$ in terms of $\alpha$ and $\beta$ magnons}
\bea
S_z({\bf k})&=& -\frac 1{\sqrt{4N}}\sum_{{\bf p,q}}\delta ({\bf k}+{\bf p}-{\bf q}) \non \\
&\Big[&\{[f_{1{\bf k}}\ell_{1\bf p}^*\ell_{1\bf q}+f_{2{\bf k}}\ell^{\prime *}_{2\bf p} \ell^{\prime }_{2\bf q}]
-[f_{3{\bf k}}m^{\prime}_{2{\bf q}}m^{\prime *}_{2{\bf p}}+f_{4{\bf k}}m_{1{\bf q}}m^{*}_{1{\bf p}}]\}
\alpha_{\bf p}^{(1)\dagger}\alpha_{\bf q}^{(1)} \non \\
&+& \{[f_{1{\bf k}}\ell_{1\bf p}^{\prime *}\ell_{1\bf q}^\prime+f_{2{\bf k}}\ell^{*}_{2\bf p} \ell_{2\bf q}]
-[f_{3{\bf k}}m_{2{\bf q}}m^{*}_{2{\bf p}}+f_{4{\bf k}}m_{1{\bf q}}^\prime m^{\prime *}_{1{\bf p}}]\}
\alpha_{\bf p}^{(2)\dagger}\alpha_{\bf q}^{(2)}\non \\
&+&\{[f_{1{\bf k}}m_{1\bf p}^*m_{1\bf q}+f_{2{\bf k}}m^{\prime *}_{2\bf p} m^{\prime }_{2\bf q}]
-[f_{3{\bf k}}\ell^{\prime}_{2{\bf q}}\ell^{\prime *}_{2{\bf p}}+f_{4{\bf k}}\ell_{1{\bf q}}\ell^{*}_{1{\bf p}}]\}
\beta_{-\bf q}^{(1)\dagger}\beta_{-\bf p}^{(1)}\non \\
&+& \{[f_{1{\bf k}}m_{1\bf p}^{\prime *}m_{1\bf q}^\prime+f_{2{\bf k}}m^{*}_{2\bf p} m_{2\bf q}]
-[f_{3{\bf k}}\ell_{2{\bf q}}\ell^{*}_{2{\bf p}}+f_{4{\bf k}}\ell_{1{\bf q}}^\prime \ell^{\prime *}_{1{\bf p}}]\}
\beta_{-\bf q}^{(2)\dagger}\beta_{-\bf p}^{(2)} \non \\
&+&\{[f_{1{\bf k}}\ell_{1\bf p}^*m_{1\bf q}+f_{2{\bf k}}\ell^{\prime *}_{2\bf p} m^{\prime }_{2\bf q}]
-[f_{3{\bf k}}\ell^{\prime}_{2{\bf q}}m^{\prime *}_{2{\bf p}}+f_{4{\bf k}}\ell_{1{\bf q}}m^{*}_{1{\bf p}}]\}
\alpha_{\bf p}^{(1)\dagger}\beta_{-\bf q}^{(1)\dagger}\non \\
&+&\{[f_{1{\bf k}}m_{1\bf p}^*\ell_{1\bf q}+f_{2{\bf k}}m^{\prime *}_{2\bf p} \ell^{\prime }_{2\bf q}]
-[f_{3{\bf k}}m^{\prime}_{2{\bf q}}\ell^{\prime *}_{2{\bf p}}+f_{4{\bf k}}m_{1{\bf q}}\ell^{*}_{1{\bf p}}]\}
\alpha_{\bf q}^{(1)}\beta_{-\bf p}^{(1)}\non \\
&+& \{[f_{1{\bf k}}\ell_{1\bf p}^{\prime *}m_{1\bf q}^\prime+f_{2{\bf k}}\ell^{*}_{2\bf p}m_{2\bf q}]
-[f_{3{\bf k}}\ell_{2{\bf q}}m^{*}_{2{\bf p}}+f_{4{\bf k}}\ell_{1{\bf q}}^\prime m^{\prime *}_{1{\bf p}}]\}
\alpha_{\bf p}^{(2)\dagger}\beta_{\bf q}^{(2)\dagger}\non \\
&+& \{[f_{1{\bf k}}m_{1\bf p}^{\prime *}\ell_{1\bf q}^\prime+f_{2{\bf k}}m^{*}_{2\bf p} \ell_{2\bf q}]
-[f_{3{\bf k}}m_{2{\bf q}}\ell^{*}_{2{\bf p}}+f_{4{\bf k}}m_{1{\bf q}}^\prime \ell^{\prime *}_{1{\bf p}}]\}
\alpha_{\bf q}^{(2)}\beta_{-\bf p}^{(2)} \non \\
&+&\{[f_{1{\bf k}}\ell_{1\bf p}^*\ell_{1\bf q}^\prime+f_{2{\bf k}}\ell^{\prime *}_{2\bf p} \ell_{2\bf q}]
-[f_{3{\bf k}}m_{2{\bf q}}m^{\prime *}_{2{\bf p}}+f_{4{\bf k}}m_{1{\bf q}}^\prime m^{*}_{1{\bf p}}]\}
\alpha_{\bf p}^{(1)\dagger}\alpha_{\bf q}^{(2)} \non \\
&+& \{[f_{1{\bf k}}\ell_{1\bf p}^{\prime *}\ell_{1\bf q}+f_{2{\bf k}}\ell^{*}_{2\bf p} \ell_{2\bf q}^\prime]
-[f_{3{\bf k}}m_{2{\bf q}}^\prime m^{*}_{2{\bf p}}+f_{4{\bf k}}m_{1{\bf q}} m^{\prime *}_{1{\bf p}}]\}
\alpha_{\bf p}^{(2)\dagger}\alpha_{\bf q}^{(1)}\non \\
&+&\{[f_{1{\bf k}}\ell_{1\bf p}^*m_{1\bf q}+f_{2{\bf k}}\ell^{\prime *}_{2\bf p} m_{2\bf q}]
-[f_{3{\bf k}}\ell_{2{\bf q}}m^{\prime *}_{2{\bf p}}+f_{4{\bf k}}\ell_{1{\bf q}}^\prime m^{*}_{1{\bf p}}]\}
\alpha_{\bf p}^{(1)\dagger}\beta_{-\bf q}^{(2)\dagger}\non \\
&+&\{[f_{1{\bf k}}m_{1\bf p}^{\prime *}\ell_{1\bf q}+f_{2{\bf k}}m^{*}_{2\bf p} \ell^{\prime }_{2\bf q}]
-[f_{3{\bf k}}m^{\prime}_{2{\bf q}}\ell^{*}_{2{\bf p}}+f_{4{\bf k}}m_{1{\bf q}}\ell^{\prime *}_{1{\bf p}}]\}
\alpha_{\bf q}^{(1)}\beta_{-\bf p}^{(2)}\non \\
&+&\{[f_{1{\bf k}}\ell_{1\bf p}^{\prime *}m_{1\bf q}+f_{2{\bf k}}\ell^{*}_{2\bf p} m^{\prime }_{2\bf q}]
-[f_{3{\bf k}}\ell^{\prime}_{2{\bf q}}m^{\prime}_{2{\bf p}}+f_{4{\bf k}}\ell_{1{\bf q}}m^{\prime *}_{1{\bf p}}]\}
\alpha_{\bf p}^{(2)\dagger}\beta_{-\bf q}^{(1)\dagger}\non \\
&+&\{[f_{1{\bf k}}m_{1\bf p}^*\ell_{1\bf q}^\prime+f_{2{\bf k}}m^{*}_{2\bf p} \ell_{2\bf q}]
-[f_{3{\bf k}}m_{2{\bf q}}\ell^{\prime *}_{2{\bf p}}+f_{4{\bf k}}m_{1{\bf q}}^\prime \ell^{*}_{1{\bf p}}]\}
\alpha_{\bf q}^{(2)}\beta_{-\bf p}^{(1)}\non \\
&+&\{[f_{1{\bf k}}m_{1\bf p}^*m_{1\bf q}^\prime+f_{2{\bf k}}m^{\prime *}_{2\bf p} m_{2\bf q}]
-[f_{3{\bf k}}\ell_{2{\bf q}}\ell^{\prime *}_{2{\bf p}}+f_{4{\bf k}}\ell_{1{\bf q}}^\prime \ell^{*}_{1{\bf p}}]\}
\beta_{-\bf p}^{(1)}\beta_{-\bf q}^{(2)\dagger}\non \\
&+& \{[f_{1{\bf k}}m_{1\bf p}^{\prime *}m_{1\bf q}+f_{2{\bf k}}m^{\prime *}_{2\bf p} m_{2\bf q}^\prime]
-[f_{3{\bf k}}\ell_{2{\bf q}}^\prime \ell^{*}_{2{\bf p}}+f_{4{\bf k}}\ell_{1{\bf q}} \ell^{\prime *}_{1{\bf p}}]\}
\beta_{-\bf p}^{(2)}\beta_{-\bf q}^{(1)\dagger}
\Big].
\eea
\bibliography{CrMWFAF}

\end{document}